# Proposed experiments to detect keV range sterile neutrinos using energy-momentum reconstruction of beta decay or K-capture events

Peter F Smith

*Department of Physics and Astronomy, UCLA, California, USA*
peterfsmith@g.ucla.edu

Sterile neutrinos in the keV mass range may constitute the galactic dark matter. Various proposed direct detection and laboratory searches are reviewed. The most promising method in the near future is complete energy-momentum reconstruction of individual beta-decay or K-capture events, using atoms suspended in a magneto-optical trap. A survey of suitable isotopes is presented, together with the measurement precision required in a typical experimental configuration. Among the most promising are the K-capture isotopes $^{131}$Cs, which requires measurement of an X-ray and several Auger electrons in addition to the atomic recoil, and $^7$Be which has only a single decay product but needs development work to achieve a trapped source. A number of background effects are discussed. It is concluded that sterile neutrinos with masses down to the 5-10 keV region would be detectable, together with relative couplings down to $10^{-10}$-$10^{-11}$ in a 1–2 year running time.

## 1 Introduction

This study investigates the feasibility of measuring the energy and momentum of each of the decay products in an atomic beta decay or electron-capture process, including the recoil atom, with sufficient precision to detect rare individual events in which a keV-range 'sterile neutrino' may be emitted. This experimental principle was originally proposed in 1992 by Cook et al [1] and by Finocchario & Shrock [2] following contemporary observations of an apparent distortion of the tritium beta spectrum suggestive of a neutrino of mass 17 keV. This idea was also foreshadowed 50 years earlier by measurements of atomic recoil energy to establish the form of the weak interaction [3].

Mass reconstruction was subsequently used by Hindi et al [5], using K-capture in $^{37}$Ar to set relative sterile neutrino coupling limits $\sim 10^{-2}$ in the mass range 370 – 640 keV, and by Trinczek et al [6] using beta decay of $^{38m}$K to set coupling limits $\sim 10^{-2}$ in the mass range 0.7 – 3.5 MeV. But for the reasons discussed below in §2, there was Increasing interest in the possible existence of sterile neutrinos at much lower masses $\sim$ 5-10 keV, and lower coupling levels $< \sim 10^{-6}$, leading Bezrukov & Shaposhnikov [6] to investigate the possibility of more precise reconstruction of tritium beta decay events, together with higher event rates, in order to reach these lower values. This was nevertheless seen as a challenging task and no experiments to reach keV neutrino masses have been attempted.

The object of the present paper is to show that a more optimistic view is now justified, and that there is both strong motivation, together with the technical capability, to construct experiments to search for sterile neutrinos in the keV mass range.

## 2 Motivation for keV mass sterile neutrinos in particle physics and astrophysics:
*(a) Particle Physics*
The motivation for higher-mass neutrinos arose originally from a theoretical belief that the existing neutrino family is incomplete, showing only left handed couplings to the intermediate bosons W and Z, in contrast with the charged leptons and the quark families which show both left and right handed couplings (Fig 2.1). This resulted in speculation that other neutrinos may exist at higher mass eigenstates with much weaker couplings to the standard flavor neutrinos, a speculation reinforced by the fact that the standard neutrinos are themselves now known to have small (sub-eV) mass eigenstates, also implying the need for both left and right handed states (for example the νMSM model [6]). This would suggest that the electron flavor neutrino emitted in beta decay will be a mixture not only of the three known mass eigenstates (via the standard mixing matrix) but would also have a small coupling $c_s$ to one (or more) sterile neutrinos, giving a correspondingly low probability of observing that sterile neutrino in a beta decay experiment. The coupling is also commonly referred to as $\sin^2 2\theta_s$ when the mixing is described by an angle $\theta_s$ [8].

$$g_L^\nu \; \bar{\nu}_e \, \gamma_\mu (1-\gamma_5) \nu_e$$
$$g_L^e \; \bar{e} \, \gamma_\mu (1-\gamma_5) e \; + \; g_R^e \; \bar{e} \, \gamma_\mu (1+\gamma_5) e$$
$$g_L^u \; \bar{u} \, \gamma_\mu (1-\gamma_5) u \; + \; g_R^u \; \bar{u} \, \gamma_\mu (1+\gamma_5) u$$
$$g_L^d \; \bar{d} \, \gamma_\mu (1-\gamma_5) d \; + \; g_R^d \; \bar{d} \, \gamma_\mu (1+\gamma_5) d$$

**Fig 2.1** Known couplings of quarks and leptons to Z boson, showing apparent absence of right handed neutrino currents (from [7]).

This small relative coupling needs to be typically $< 10^{-3} - 10^{-4}$ to be consistent with existing particle physics limits [4] summarized in Fig 2.2.

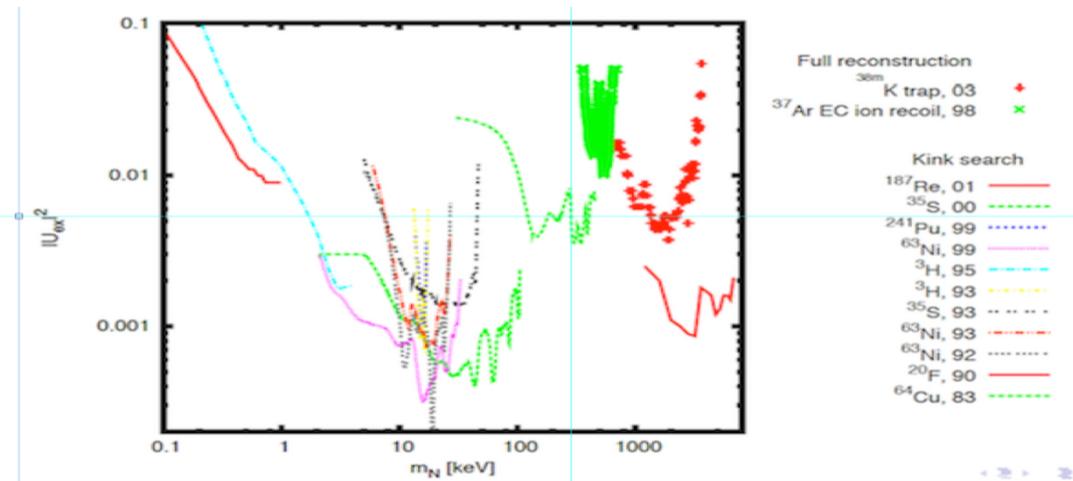

**Fig 2.2** (taken from [9]) Limits on mass and relative coupling of 'sterile' neutrinos from terrestrial experiments

*(b) Dark Matter*
An important astrophysical consequence of the possible existence of quasi-sterile neutrinos is that they have become a viable candidate for the as-yet unidentified dark matter in the universe, notably the non-luminous material that gravitationally dominates the mass of our own galaxy. During the past 30 years, identifying the dark matter has become a priority cosmological problem, with extensive effort and funding devoted to increasingly large and more sensitive underground detectors, mainly devoted to searches for hypothetical 5 – 500 GeV range weakly interacting massive particles (WIMPs), in particular for the lightest neutral particle predicted by supersymmetry theory. However, at the time of writing these searches continue to give null results, and, along with absence, so far, of evidence for supersymmetry at the Large Hadron Collider, this has given rise to increasing interest in alternative explanations for the dark matter, including the 'sterile' neutrino. The latter could decay into a standard neutrino (via an intermediate W and lepton) but a sufficiently small coupling $\leq 10^{-4}$ leads to a lifetime for this process exceeding that of the universe by many orders of magnitude [10][11][12] for a neutrino mass in the keV range.

Sterile neutrino masses as large as 1-10 keV were first predicted by Shi and Fuller [12] in 1999, on the basis of production mechanisms that could account for the present dark matter density ~ 0.4 GeV cm$^{-3}$. This was confirmed by further theoretical work in the ensuing 10-15 years [13 – 20]. Thus sterile neutrinos could provide a background of non-luminous matter that is essentially stable on cosmological timescales. Masses somewhat higher than 1-10 keV might be possible with different cosmological assumptions, but below this range there is a lower mass limit of 0.4 keV imposed by applying the standard Tremaine & Gunn [21] phase space limit to the dark matter density in dwarf galaxies. The conclusion is that one or more sterile neutrinos in the keV mass region above would be a plausible candidate for the Galactic dark matter.

*(c) Possible astrophysical X-ray observation*
If quasi-sterile neutrinos constitute all or part of dark matter in galaxies, their slow decay into ordinary neutrinos via intermediate bosons and leptons (Fig 2.3) would result in X-ray lines of energy half (due to momentum conservation) of the sterile neutrino mass. These would need to be strong enough to be seen above the existing X-ray background. Analysis of X-ray data from several sources of dark matter, including galactic clusters [17] [18], provides an approximate limit contour of coupling versus neutrino mass (shaded region in Fig 2.4).

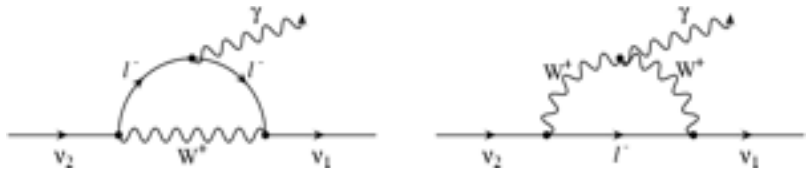

**Fig 2.3** Two diagrams for decay of sterile neutrino to standard neutrino [8] [13]

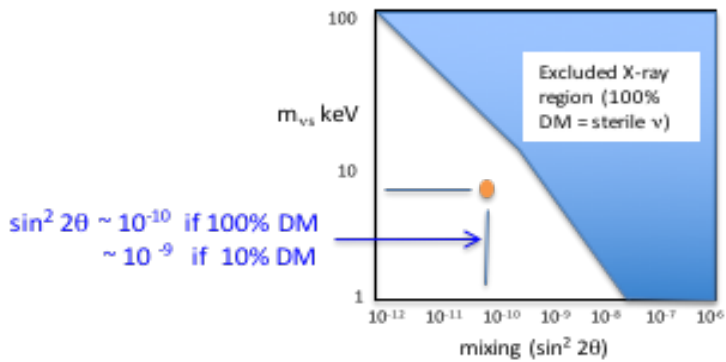

**Fig 2.4** Approximate excluded range of sterile neutrino mass and coupling, assuming the dark matter to be 100% sterile neutrinos. The single orange point shows the approximate mass and coupling for a sterile neutrino interpretation of an unexplained astronomical X-ray peak.

The 'exclusion curve' is only approximate since it depends on assumptions about the dark matter distribution in the astronomical sources and the % of dark matter (up to 100%) attributable to sterile neutrinos. But outside the excluded region there have been recent observations of an unexplained 3.6 keV X-ray line, from both galaxy clusters and our own Galactic center [24-26] which, if due to a sterile neutrino decay, would correspond to a mass 7.2 keV and coupling ~$10^{-10}$, consistent with the range of theoretical expectations. However, the overall case for sterile neutrinos, outlined in (a) and (b) above, remains valid even if this apparent signal proves to have another explanation.

## 3 Detection of keV-range sterile neutrinos

There are two types of possible detection methods:
(a) If the galactic dark matter consists of sterile neutrinos, then direct detection may be possible
(b) Whether or not sterile neutrinos form the dark matter, experiments involving weak decay could detect rare events with emission of sterile neutrinos.

This paper is concerned principally with laboratory experiments based on method (b) but we first summarize published work on method (a). No existing dark matter detectors would be sensitive to sterile neutrino dark matter, because the nuclear recoils would be too low in energy (~ $10^{-2}$ eV), so new methods are needed.

In principle, direct detection would be possible by coherent scattering from atomic electrons, by the two processes shown in Fig 3.1. This has been investigated by Ando & Kusenko [27], showing that the scattering would transfer sufficient energy to produce ionization in a detector material, but the rate (assuming zero background) would be only 1 event/year/kiloton for a sterile neutrino mass $m_s$ = 7 keV and coupling $c_s = 10^{-9}$, with proportionality to $c_s$ and $m_s^2$.

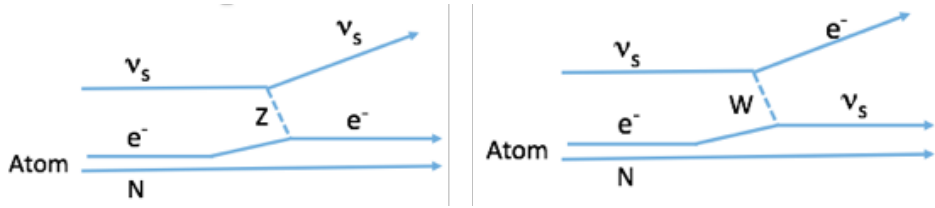

**Fig** 3.1 Processes for detection of dark matter sterile neutrinos by electron scattering

The decay processes producing X-rays (Fig 2.3) could in principle be seen in a terrestrial experiment consisting of a large evacuated volume with its surface covered with low background X-ray detectors. But the required detector volume and area would be impractically large. Assuming a dark matter density 0.3 GeV/cm$^3$, $c_s = 10^{-9}$ and $m_s$ = 7 keV, a km$^3$ volume with its 6 km$^2$ surface fully covered with zero background detector less than ten 3.5 keV X-rays would be produced per year, the number in this case being proportional to $m_s^5$ [27].

A further theoretical possibility, using macroscopic coherence, is that of coherent scattering from an array of sub-micron granules suspended in an 'optical honeycomb'. This currently lies beyond existing technology, but remains a future possibility with increasing proficiency in nanotechnology [28] [29].

An apparently more promising method of direct detection would be by inverse beta decay. Any beta decay process A -> B + e⁻+ (anti)ν can be inverted to ν + A -> B + e⁻, giving events beyond the normal beta spectrum end point, and fully separated from the latter by the neutrino mass. An .example for Tritium decay is illustrated in Fig 3.2. This method was discussed over 30 years ago by Irvine & Humphries [30] as a possible means of detecting the relic neutrino background, but requiring a target of at least 1 g tritium (subsequently corrected to 100g tritium [31]), many orders of magnitude greater than the targets of a few μg typically in use at that time. Thus detecting the relic neutrinos in this way seemed out of reach in the 1980s. Now however, thirty years later, the PTOLEMY project [32] is ambitiously planning to construct the required 100 g tritium target as a single atomic layer on the surface of an array of graphene nanotubes of total area $10^4$ m$^2$, a reminder that an apparently unfeasible idea can sometimes become feasible within a generation.

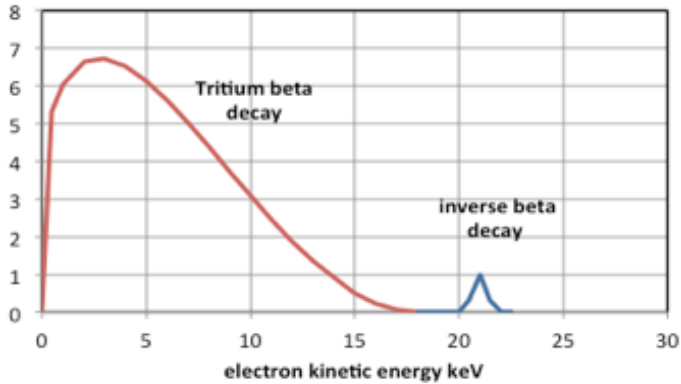

**Fig 3.2**  Illustration of the detection of sterile neutrino dark matter by inverse beta decay in a tritium target. The peak is displaced from the spectrum end point by the sterile neutrino mass

Application of inverse beta decay to sterile neutrino dark matter detection has been considered by a number of authors [33 - 40] estimating the target mass needed to obtain a significant peak of events beyond the beta spectrum end point. Li & Xing [34] find the optimum isotopes to be tritium $^3$H (lifetime ~ 12 years, beta end point 18 keV) and $^{106}$Ru (lifetime ~ 1 year, beta end point 39 keV).  To obtain 10 events/year would require target masses 10 kg $^3$H or 1000 kg $^{106}$Ru, the latter of course needing annual replacement to maintain the decay rate.  The targets would need to be in the form of a gas or gaseous compound or thin surface layers to allow the electrons to escape to a (low background) detector.  These formidable requirements are not feasible in the near future, but, by analogy with the factor $10^7$ generational increase in tritium target mass mentioned above, remain a future possibility for direct sterile neutrino dark matter detection.

Moving from direct detection to laboratory detection, the most familiar existing type of search for sterile neutrinos in the range 1 – 1000 keV is that of a possible distortion of the electron spectrum from beta decay of various isotopes.  This could result from a small admixture of events involving the emission of a higher mass neutrino, giving a slight distortion of the electron spectrum (a 'kink') at a distance from the end point approximately given by the sterile neutrino mass, as illustrated in Fig 3.3.  The latter exaggerates the effect for illustration, and to change the spectrum by a measurably significant amount, the fraction of decays giving a higher mass neutrino in general needs to be > 0.1%, giving rise to the limits shown in Fig 2.2 for a number of experiments (references for these listed in [41]).

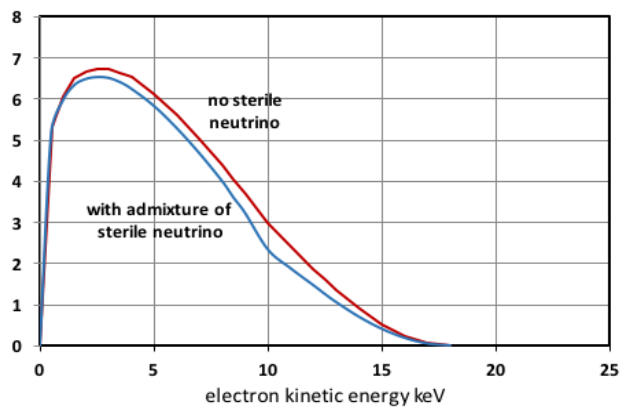

**Fig 3.3**  Illustrative effect of admixture of a ~ 7 keV sterile neutrino population on the normal ( ~ zero neutrino mass) tritium beta spectrum

To obtain sensitivity to smaller values of the mixing parameter, a stronger beta source is needed, giving orders of magnitude more events/year, along with a higher energy resolution to search for significant distortions in the spectrum. The largest and most sensitive such experiment, currently under construction, is the 10m diameter KATRIN spectrometer [42] which is designed to achieve an energy resolution of 0.2 eV, with the objective of measuring, or placing limits on, the largest of the mass eigenstates from the end point of the spectrum. Detailed analysis of the whole spectrum shape will also achieve sensitivity to any sterile neutrinos with mass < 18 keV [43, 44]. Fig 3.4, taken from [44] summarizes the expected sensitivity to sterile neutrino masses and couplings, for two different analysis techniques. However, this depends on the non-trivial development of a detection system capable of a counting rate ~ $10^{10}$ counts per second, in order to analyze the whole spectrum, rather than just the end point [44]. If this is possible, together with the currently planned (2016) tritium source luminosity, $10^{11}$ decays/s, a 90% confidence exclusion limit $\sin^2\theta_s \sim 10^{-7}$–$10^{-8}$, for masses 1 – 17 keV, could be achieved in 3 years active running time, and ~ $10^{-7}$ for masses 5 – 15 keV for a reduced luminosity of $10^9$ decays/s (if necessary to reduce systematic effects [44]). There are future prospects for increasing the source luminosity by a factor 100, giving a further order of magnitude reduction in coupling sensitivity (again subject to an upgraded detection system allowing a correspondingly increased counting rate).

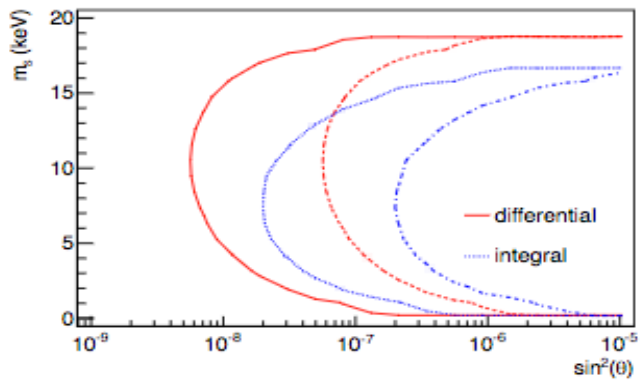

**Fig 3.4** (taken from S Mertens et al [44]). Expected sterile neutrino coupling sensitivity for KATRIN experiment for a 3-year active running time, for two different analysis procedures (red and blue full curves) with the planned strength of $10^{11}$ decays/s and an assumed counting rate increased to 1010 cps. The dashed curves show the corresponding limits for a source strength $10^9$ decays/s.

It is clear that in preference to attempting to detect a small sterile neutrino event population within a much larger population of standard neutrino events, it would be preferable to isolate the individual sterile neutrino events as a separate population. This is the objective of the principle of full energy-momentum reconstruction already outlined in §1 above and the subject of references [1] – [6]. Two distinct variations of this are shown in Fig 3.5. In Fig 3.5(a) a free electron is emitted with a continuous (kinetic) energy spectrum $E_{eK}$, together with an antineutrino also with a continuous energy spectrum. In Fig 3.5(b) an electron is captured from an atomic shell, usually the K-shell, together with a mono-energetic neutrino. In each case, the necessary momentum precision requires measurement by time of flight, in turn necessitating a trigger event to define the moment of decay. In case (a) an isotope needs to be chosen that emits a simultaneous nuclear gamma, of known energy-momentum but whose emission direction must be measured. In case (b), the K-capture is followed within ~ 1ps by the filling of the atomic vacancy by emission of an X-ray, also of known energy-momentum, and also needing measurement of its emission direction. Although K-capture appears to have the considerable advantage of not requiring momentum measurement of the electron, the filling of the K vacancy leaves another unfilled vacancy in a higher shell, and the resulting rearrangement of atomic states results in the emission of several Auger electrons within a further ~ $10^{-13}$ - $10^{-15}$ s. These are of relatively low kinetic energy, but turn out to be significant for the mass reconstruction and all need to be collected and measured. These two cases will be discussed separately in sections §5 and §6 below.

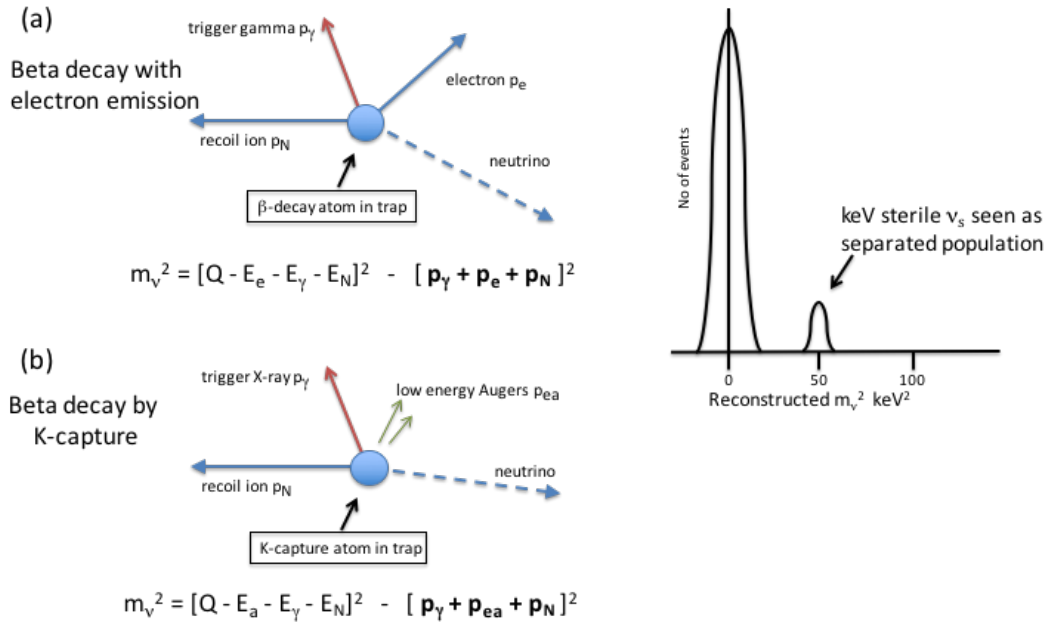

**Fig 3.5**   Principle of full energy momentum reconstruction in beta decay. Precise measurement of vector momentum of all observable particles (including the atomic recoil) together with the known decay energy release Q, allows the energy and momentum of the unseen neutrino to be calculated, and hence its mass from the formula shown. Measurement errors will spread the values around a mean value, but the majority of events form a peak close to zero mass, while any higher mass neutrinos will form, for sufficiently small momentum resolution, a separate peak. Two distinct types of beta decay are shown, in each case needing an additional photon to provide a time origin for the time of flight momentum measurements: (a) with electron emission, using an isotope with an additional nuclear gamma as trigger. (b) by absorption of an electron from an atomic shell – usually K-shell, then releasing an atomic X-ray as a trigger.

## 4  Available measurement techniques

Before discussing the neutrino mass range achievable by energy-momentum reconstruction, it is necessary to discuss the measurement precision attainable with available techniques. Since measurement of both momentum and direction are necessary, the former requires accurate timing over a suitable path length, while the latter requires also position sensitivity to define the 3-D emission directions of all the particles. The ability to do the latter results from the development over the past 30 years of the COLTRIMS technique (Cold Target Recoil Ion Momentum Spectroscopy) illustrated in Fig 4.1

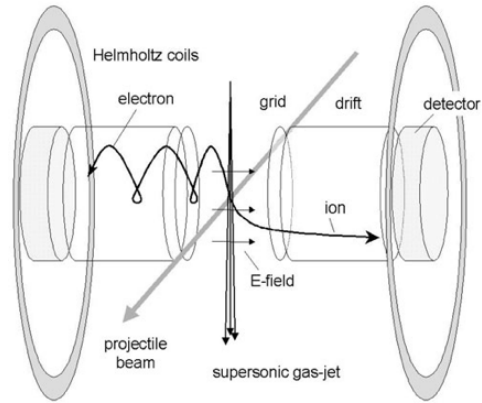

**Fig 4.**1  General arrangement of components of a COLTRIMS system (from [46])

Full details and examples of this technique are given in reviews [45] and [46]. In the case illustrated, the reaction products emitted from a small volume, defined by the interaction of two atomic beams in high vacuum, are directed on to position-sensitive detectors, using magnetic and electric fields to increase the solid angle captured. In the case of the recoil ion, the detector is an MCP (Multi-Channel Plate) preceded by a ~2kV grid to accelerate the ion (which has a very low recoil kinetic energy) on to the MCP surface. An MCP can also be used for the electron detection and emission angle, the principal alternative being a traditional multi-anode photomultiplier tube, or a multi-anode silicon photomultiplier (SiPM) or a Multi-Pixel Photon Counter (MMPC). For the neutrino mass sensitivity figures calculated in this paper, the position and timing resolution achievable with commercially available components and systems is assumed [47] – i.e. a typical timing resolution ~ 0.2 ns rms (hence a total time of flight resolution ~ 0.3 ns), and a typical position resolution ~ 20 μm.

In the case of the decay products from a single atomic species, it would be possible in principle to define the source by exposing just a short length of an atomic beam, but a preferable solution has been to create a small stable volume of atoms in a Magneto-Optical Trap (MOT). An MOT confines and cools the atoms at the intersection of three retro-reflected laser beams and a quadrupole magnetic field provided by a pair of anti-Helmholtz coils (Fig 4.2). Experiments with trapped atoms have been reviewed by Behr [48] and by Melconian [49]. If such a trap is used as source for a configuration analogous to COLTRIMS, the technique is often referred to as MOTRIMS [50-53].

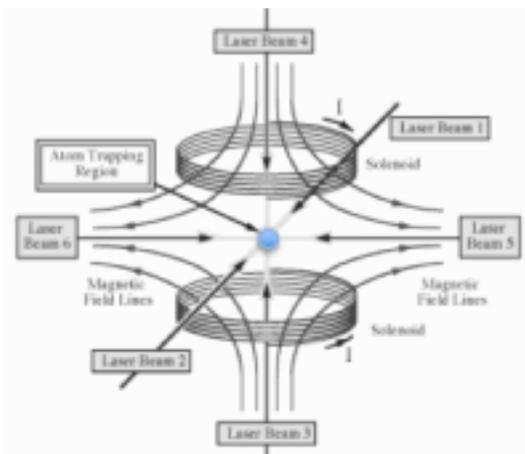

**Fig 4.2**  (from [53]. Schematic arrangement of magnetic fields and three retro-reflected laser beams,
to produce a spherical trapped cloud of cold atoms in high vacuum

From the viewpoint of sterile neutrino searches, there are several important interlinked parameters for the cloud of trapped atoms:

(a) The effective temperature T should sufficiently low to avoid the initial atom velocity contributing a significant time of flight error (typically required to be < $10^{-4}$). Values of T ~ 25 - 100 μK or below are possible [54] corresponding, for an atom with A ~ 100, to an rms velocity ~ 5 – 10 cm/s – small compared with typical recoil velocities ~ $10^5$ -$10^6$ cm/s, but nevertheless can have a significant effect on the rms spread of the reconstructed distribution of $m_s^2$.

(b) Similarly the rms radius σ of the atom cloud needs to be << L, the chosen recoil path length, but because the practical maximum peak trapped atom density $N_0$ is typically no higher than ~3 - $5.10^8$ atoms/mm$^3$ [54], σ will increase with the total number $N_T$ of trapped atoms, and assuming a three-dimensional Gaussian distribution, approximately as:

$$\sigma_3 \text{ (mm)} \sim [N_T/ 10^9]^{0.33} \qquad (4.1a)$$

showing that $\sigma_3$ can range typically from ~ 0.1mm at $N_T = 10^6$ to ~ 2mm at $N_T = 10^{10}$.

When viewed from a specific direction, appropriate integration gives:

$$\text{for a two-dimensional projection} \quad \sigma_2 \text{ (mm)} \sim [N_T/ 10^9]^{0.33} \times [2/3]^{0.5} \qquad (4.1b)$$
$$\text{for a one-dimensional projection} \quad \sigma_1 \text{ (mm)} \sim [N_T/ 10^9]^{0.33} \times [1/3]^{0.5} \qquad (4.1c)$$

Eq.(4.1b) will be appropriate when considering source errors in measured emission angles.

To be sensitive to low values of the sterile neutrino coupling, the total number of trapped atoms needs to be as large as possible to maximize the number of decays per year. Camara et al [55] achieved an MOT loading of $N_T = 1.3.10^{11}$ Rb atoms, finding a somewhat larger exponent, ~ 0.35 - 0.39 in eq.[4.1].

(c) For an isotope with a half life $\tau_{1/2}$(days) the decay rate per atom is $\lambda = 0.69/\tau_{1/2}$ per day and hence a source decay rate

$$dN/dt = 0.69 \, N_T/\tau_{1/2} \text{ per day}$$
$$= 250 \, N_T/\tau_{1/2} \text{ decays per year (of active running time)} \qquad (4.2)$$

Thus isotopes with low values of $\tau_{1/2}$ (eg 3-100 days) are advantageous, though requiring frequent replenishment. For traps fed continuously from an accelerator beam, shorter half lives (< 1 day) become usable.

(d) Although the recoil momentum could be calculated simply from the time of flight $t_F$ between the trigger moment and the MCP detection time, over a known path length L, it is usual to add an axial electric field of a few V/cm to extract the ions from the source region. This actually has four advantageous purposes:

(i) To increase the solid angle for capture of the recoil ions, through the focusing effect of the axial electric field.

(ii) To provide a controllable proportionality between recoil momentum differences and time of flight differences. Since the recoil is non-relativistic, simple kinematics gives (to a sufficiently high approximation) the following linear relationship between momentum differences $\Delta p_N$ (keV/c), time of flight differences $\Delta t_F$ (s) and a (uniform) electric field F (kV/cm) [45][46]:

$$\Delta p_N = c.n.F.\Delta t_F \qquad (4.3)$$

where $c = 3.10^{10}$ cm/s and F = V(kV)/L (cm), n = ion charge multiplicity

(iii) To provide compensation for the time of flight differences between decays from the front and back of the significant diameter source atom cloud. This is achieved by providing a uniform electric field over a distance $L_1$ from the source, followed by a field free region of length $L_2 = 2L_1$ [55][56]. The total flight time is then to a first approximation independent of small changes in $L_1$ arising from the finite source diameter. A more exact criterion, for field length $L_1$, field free length $L_2$, initial recoil velocity u and final velocity v, is:

$$L_1/L_2 = (v^2 - u^2)/2v^2 \qquad (4.4)$$

So that the factor 2 ratio applies if u << v, otherwise [4.4] can be used.
This longitudinal compensation is referred to as "McLaren focusing" or "Time focusing". Transverse position and momentum focusing configurations have also been studied [45][46][56] in order to maximize momentum resolution of the system.

(iv) To capture ions emitted in both the facing and back hemispheres (Fig 4.3). For each there is a computable unique correspondence between the 2D detection co-ordinates and the polar and azimuthal emission angles, but the correspondence differs for front and back hemispheres. This gives a two-fold ambiguity in the emission angles, which can subsequently be resolved in the mathematical reconstruction of the event, one of the two alternatives giving an unrealistic value of $m_s^2$.

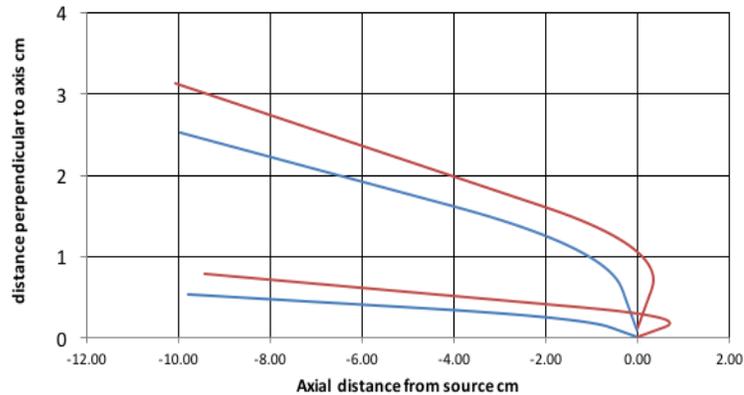

**Fig 4.3** Typical computed ion trajectories with axial electric field ~ 2V/cm, showing collection of recoils from both hemispheres. The MCP is located on the left hand side, with paths emitted towards the MCP in blue, paths emitted away from MCP shown in red.

## 5 Choice of Isotopes with a beta spectrum

Fig 3.5 shows the two distinct types of nuclear beta decay available for a sterile neutrino search (a) emitting a free electron, sometimes with a nuclear gamma, and (b) capturing the electron in the nucleus followed by emission of an atomic X-ray and low energy Auger electrons. Of type (a), tritium has been adopted for measuring the mass of standard neutrinos, and as discussed in §3 has also been suggested for sterile neutrino searches. However, this does not provide a clear trigger to define t = 0 for time of flight measurements, and would not detect a sterile neutrino with a mass larger than its 18 keV end point. This section considers the optimum choice of type (a) isotopes with a clear trigger and a wider mass range. Numerical details of Isotope decay schemes were obtained from the standard tables of Firestone and Shirley [58].

There are 147 beta decay isotopes with half-lives between ~ 3 days and 2 years. 65 of these decay by electron capture and are considered in §6. The remaining 82 candidates were reduced by the requirement of a single prompt ( ~ ps) gamma for use as a trigger, and rejecting the following:

  (a) positron decays.
  (b) decays to the nuclear ground state giving no gamma.
  (c) delayed gammas.
  (d) nuclear cascade giving multiple gammas.
  (e) complex sets of several beta decays and gammas.
  (f) gamma probability < 0.1%

This left 17 candidates, listed in Table 5.1 in order of decay rate (per atom per year) obtained by multiplying the intrinsic decay rate eq.[4.2] by the fraction of decays producing the trigger gamma. Table 5.1 shows also the beta end point energy, gamma energy, and Q value for the decay.

Table 5.1  Beta decay isotopes with continuous electron spectrum, giving a single prompt gamma trigger, listed in order of triggered decay rate/year/atom (for number in MOT maintained constant).

| Z | A |  | half life days d | total Q keV | trigger gamma keV | fraction gamma | beta end point keV | Trigger rate per year per atom | MOT |
|---|---|---|---|---|---|---|---|---|---|
| 79 | Au | 198 | 2.7 | 1372 | 412 | 0.98 | 960 | 91.4 | |
| 53 | I | 131 | 8.0 | 971 | 361 | 0.81 | 610 | 25.5 | |
| 79 | Au | 199 | 3.1 | 453 | 208 | 0.21 | 245 | 17.1 | |
| 88 | Ra | 225 | 14.9 | 356 | 40 | 0.53 | 316 | 9.0 | <-- |
| 71 | Lu | 177 | 6.7 | 498 | 113 | 0.20 | 385 | 7.5 | |
| 41 | Nb | 95 | 35.0 | 925 | 765 | 1.00 | 160 | 7.2 | |
| 60 | Nd | 147 | 11.0 | 896 | 92 | 0.28 | 804 | 6.4 | |
| 80 | Hg | 203 | 46.6 | 492 | 280 | 0.81 | 212 | 4.4 | <-- |
| 40 | Zr | 95 | 64.0 | 1124 | 758 | 0.98 | 366 | 3.9 | |
| 58 | Ce | 141 | 32.5 | 580 | 145 | 0.48 | 435 | 3.7 | |
| 79 | Au | 196 | 6.2 | 686 | 426 | 0.08 | 260 | 3.2 | |
| 26 | Fe | 59 | 44.5 | 1565 | 1099/1291 | 0.43 | 476/274 | 2.4 | |
| 70 | Yb | 175 | 4.2 | 470 | 114 | 0.02 | 356 | 1.1 | <-- |
| 58 | Ce | 144 | 285 | 319 | 133/80 | 0.23 | 186/239 | 0.20 | |
| 69 | Tm | 170 | 129 | 968 | 84 | 0.03 | 884 | 0.06 | <-- |
| 68 | Er | 169 | 9.4 | 351 | 8 | 0.002 | 343 | 0.05 | <-- |
| 45 | Rh | 102 | 207 | 1150 | 556 | 0.02 | 594 | 0.02 | |

Not all of these are suitable for trapping in an MOT. Aside from hydrogen and the inert gases, the elements so far known to be feasible for trapping in an MOT are listed in Table 5.2. review by Behr & Gwinner [48] lists 15 elements that had been successfully trapped at that time:

Table 5.2   Elements suitable for Magneto-optical trap

| From review by Behr  & Gwinner, 2009 [48] | Additional elements added subsequently [54] |
|---|---|
| Li, Na, Mg, K,  Ca, Cr, Rb, Sr, Ag, Cs, Ba, Er, Yb, Hg, Ra | Dy, Tm, Cd |

Some of these, Yb, Ag, Dy, Er, Tm, have been loaded into an MOT more efficiently via a buffer gas cell [59] which has also been used for many previously difficult compounds and radicals and may allow more elements to be considered for MOT sources. However, for the present purpose only the those listed in Table 5.2, already demonstrated to be feasible, will be considered. This in turn means that only five of the elements in Table 5.1 are suitable for an MOT source, as shown in the final column. Of those, three are in the lower part of the list, and thus less favorable from the viewpoint of event rate, while [225]Ra has the disadvantage of a more difficult supply route via thorium [54]. This leaves [203]Hg, which has a 280 kev trigger gamma, a 212 keV beta end point, a Q of 492 keV, and well-suited for loading into an MOT with high efficiency [54]. Thus $^{203}$Hg provides an ideal example for assessing the feasibility of reaching keV-range sterile neutrino masses, through the decay process of Fig 3.5(a). The latter shows the reconstruction of the neutrino mass via the relation $m^2$ = (total visible energy)$^2$ – (net visible momentum)$^2$, written in terms of the observable measurable quantities (using units in which energy, momentum, and mass are all in keV):

$$m_\nu^2 = [Q - E_\gamma - E_{eK} - E_N]^2 - [\mathbf{p}_\gamma + \mathbf{p}_e + \mathbf{p}_N]^2 \quad (5.1)$$

(where the first bracket consists of scalars and the second bracket vectors)

Since the emitted gamma energy/momentum is known, eq.(5.1) requires the measurement of the other two momenta by time of flight, and the three momentum directions by position-sensitive detectors. Differentiation of eq.(5.1) enables the rms spread of $m^2$ to be directly related to the rms spread of any of the measured parameters, and hence to the required measurement precision needed to achieve sensitivity to a given value of neutrino mass. Beginning with the time of flight measurements, and using the approximate equality of the two terms on the right hand side of eq.(5.1):

$$d(m_\nu^2) = 2[Q - E_\gamma - E_{eK} - E_N]\, d\mathbf{p}_N$$

$$= 2[210\text{ keV} - E_{eK}]\, d\mathbf{p}_N \quad (5.2a)$$

From Fig 5.1, 80% of the electron spectrum lies in the range ~ 20 - 140 kev, averaging ~ 80 keV. Thus

$$d(m_\nu^2)\ (\text{keV}^2) \approx 260\ (\text{keV}) \times d\mathbf{p}_N\ (\text{keV}) \quad (5.2b)$$

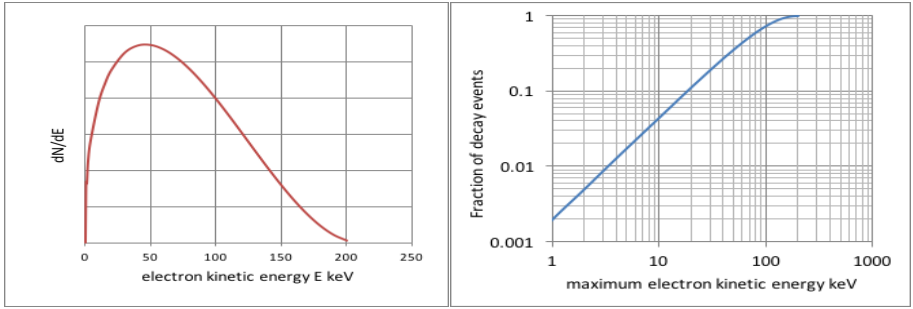

**Fig 5.1** (left) shape of differential beta spectrum for $^{203}$Hg, with Fermi correction [60]
(right) cumulative fraction of decays versus maximum electron kinetic energy

This can be converted to the time of flight precision needed to achieve a given $d(m_\nu^2)$, by noting that the atomic recoil is non-relativistic so that $p_N$ (keV)= $M_N$ (keV)$v_N$/c with $v_N$ = $L_N/t_N$, where $t_N$ (seconds) is the travel time for a path length $L_N$ (cm). This gives:

$$d(m_\nu^2)\ (\text{keV}^2) \approx 260\ (\text{keV}) \times M_N\ (\text{keV}) \times (v_N/c)^2 \times (c\, dt_N/L_N) \quad (5.3)$$

For the average (and most probable) opening angle between electron and neutrino ~ 90° [61] and isotropic gamma emission, the average recoil momentum is 320 keV, with $v_N/c$ ~ 1.6x10$^{-6}$, giving:

$$d(m_\nu^2)\ (\text{keV}^2) \approx 4 \times dt_N\ (\text{ns})/L_N\ (\text{cm}) \quad (5.4)$$

This immediately shows that there would be no problem in obtaining the required precision for the momentum of the recoiling atom, since a time-of-flight measurement accuracy ~ 10ns and a path length ≥ 50 cm will give a spread < ±1 keV$^2$ in the reconstructed (mass)$^2$. However, the electron presents a more severe problem because of its much higher velocity over most of its kinetic energy spectrum.

Differentiating eq.(5.1) with respect to pe, this time with both RHS terms contributing, eqs.(5.3) and (5.4) become, for an average electron energy $E_{eK} \approx$ 80 keV, velocity $v_e/c \approx$ 0.50 and path length $L_e$:

$$d(m_\nu^2) \approx 2\,[\,210 \text{ keV} - E_{eK}\,]\,[\,1 + v_e/c\,]\,d\mathbf{p_e} \text{ (keV)} \quad (5.5)$$

with $\quad dp_e \approx m_e\,(v_e/c)^2\,c\,dt_e\,/L_e$

giving $\quad d(m_\nu^2)\,(keV^2) \approx 1.5.10^6 \times dt_e(ns)\,/L_e\,(cm) \quad (5.6)$

In this case, even assuming (in a search for a sterile neutrino mass of 7 keV) a (mass)$^2$ precision of (50 keV)$^2 \pm$ 20 (keV)$^2$ and a minimum timing error of 0.3 ns, the required time-of-flight path would be $L_e >$ 200 metres. Assuming this to be impractical, at least for initial experiments, the path length can be reduced to a more practical value if only the lowest 0.5% of the electron spectrum is used, i.e. up to an energy ~ 2.5 keV, with an average $v_e/c \approx$ 0.05. Eq.(5.6) then becomes:

$$d(m_\nu^2)\,(keV^2) \approx 1.6.10^4 \times dt_N(ns)\,/L_e\,(cm) \quad (5.7)$$

Thus using only the lowest 0.5% of the beta spectrum would reduce the required path length to ~ 2 m, at the expense of factor 200 loss in number of decays processed.

But even assuming this to be acceptable, or the search confined to limits on larger sterile neutrino masses, the problem of measuring the three momentum angular directions needs to be assessed.

The right hand side of eq.(5.1) contains three vector products of the form $2p_x.p_y.\cos\theta_{x,y}$. As before, differentiation gives a relation between d(mass)$^2$ and an angular measurement error d(cosθ).

$$d(m_\nu^2)\,(keV^2) \approx 2\,p_x.p_y.\sin\theta_{x,y}\,d\theta_{x,y} \quad \text{(for each of the angles } \theta_{N,e},\,\theta_{e,\gamma},\,\theta_{N,\gamma}) \quad (5.8)$$

Using average values $p_{Nav}$ ~ 320 keV, $p_{eav}$ ~ 80 keV, $p_{\gamma av}$ = 280 keV, $(\sin\theta_{x,y})_{av} \approx$ 0.64:

$$d\theta_{N,e} \approx 2.10^{-5}\,d(m_\nu^2) \quad d\theta_{e,\gamma} \approx 2.10^{-5}\,d(m_\nu^2) \quad d\theta_{N,\gamma} \approx 5.10^{-6}\,d(m_\nu^2) \quad (5.9)$$

In terms of detection position errors ds at a distance R, dθ = ds/R (radians), so for example, at R = 100cm, a mass$^2$ error ~ 10 keV$^2$ requires position errors ~ 0.1mm for detection of both recoil atom and electron for a determination of $\theta_{N,e}$. This is achievable with the high resolution (< 40 micron) structure of an MCP detector surface, but a significant problem for the 280 keV gamma for which the detection by an MCP has only 1-2% efficiency.

A better solution would be to use scintillating plates followed by silicon photomultipliers (SiPM) with 50 micron pixels [62]. Since sub-ns timing is also required, the scintillator needs to be either LYSO, which has a 47 ns decay but a high light output so provides a few photoelectrons < 1 ns, or BaF$_2$ which provides both a fast (0.6 ns) light output pulse for timing plus a larger but slower (600ns) light output pulse for position measurement [63]. For BaF$_2$ the fast pulse is in the UV wavelength (220 nm) necessitating coating the SiPM with TPB wavelength shifter to convert to a wavelength suitable for the SiPM [64]. However, for a 280 keV gamma from $^{203}$Hg. the BaF$_2$ scintillator needs to be 6 mm thick to obtain even a 50% detection efficiency, thus giving a transverse spread of photons of this order. If the thickness is reduced to 1mm to allow sub-mm interpolation of the photon signals, the gamma detection efficiency drops to 10% - giving another order of magnitude loss, in addition to the two orders of magnitude loss in processed decay events arising from the above limitation to the lowest < 1% of the beta energy spectrum. For these reasons LYSO may be the preferred overall choice for an $^{203}$Hg trigger.

A larger fraction of the electron spectrum may be possible by applying a longitudinal magnetic field to give the electrons a spiral path to shorten the geometric length of the system. From Fig 5.1, 10% of the decays have energies up to 20 keV, with an average $(v_e/c) \sim 0.14$. This increases the numerical coefficient in eq.(5.8) to

$$d(m_\nu^2) \text{ (keV}^2) \approx 1.2 \cdot 10^5 \times dt_N(ns) / L_e \text{ (cm)} \qquad (5.10)$$

giving a required required path length to $L_e \sim 18$m. Suppose now that an axial magnetic field of 16 Gauss is applied giving, for 20 keV electrons emitted at an angle $\alpha > 80°$ to the axis ($\cos \alpha \sim 0.1$) a spiral path of radius ~30cm (Fig 5.2) and circumference ~1.8 m. Since the ratio of axial/transverse speeds is ~1/10, ten spiral turns will give the required path length of 18 m, reaching the detector at an axial distance now shortened to ~1.8 m. However, for smaller emission angles the axial/transverse speed is greater, giving fewer spiral turns, a smaller spiral path and a larger $d(m^2)$ error from eq.(5.10).

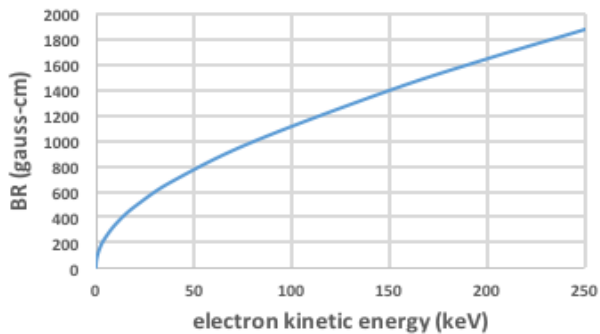

**Fig 5.2** Product of magnetic field B and radius of curvature R, versus electron kinetic energy

So the possibility of shortening the geometric length of the system with an axial magnetic field applies, in this example, to only 10% of the solid angle. The magnetic field enables a larger solid angle to be collected and focused onto a detector at a given distance, but does not significantly improve the time-of-flight precision for most of the events. Moreover, since the emission angle is calculated from the radial co-ordinate r of the detection point [46], a significant fraction of events for which the number of spiral turns is within ± 0.1 of any integer, and hence returning to points close to r ~ 0, cannot be utilized.

A background effect arising from the use of a gamma trigger would be the interaction of the nuclear gamma with atomic electrons, ejecting an electron by internal conversion or Compton effects. This also leaves a vacancy giving rise to atomic rearrangement and Auger electron emission. From the formula in [65] this effect should be $< 10^{-3}$ of the decays, with these events sufficiently distinctive to be rejected.

Some of the preceding problems might be alleviated by replacing $^{203}$Hg with an isotope with a lower energy gamma trigger, having a smaller kinematic contribution, in particular Tm, Yb, and Er in Table 5.1. These are also suitable for an MOT, but each has a compensating disadvantage of a low fraction of decays giving the gamma trigger, and also a higher energy beta end point. Thus all of the candidate beta isotopes, although in principle able to reach reconstructed masses < 10 keV, have a low overall efficiency of usable decays, typically $< 10^{-3} - 10^{-4}$ making it difficult to reach very low sterile neutrino coupling levels, e.g. $10^{-6}$ and possibly down to $10^{-10}$ (Fig 2.4), even with an MOT filling ~ $10^8 - 10^{10}$ atoms.

Isotopes decaying by K-capture have the apparent advantage of being a simpler two-body decay, with the electron absorbed rather than emitted, but the disadvantage of releasing several low energy Auger electrons. These are considered in the next section.

## 6 Choice of Isotopes decaying by electron capture

Electron capture isotopes have the basic advantage of an initial two-body decay releasing a mono-energetic neutrino, and a recoil atom, with equal and opposite momenta. Since the kinetic energy of the recoil is much smaller than the energy of the neutrino, emission of a sterile neutrino with a larger mass, will have a lower momentum, resulting in a slightly lower momentum for the recoil atom. The majority of electron captures are from the K shell, leaving a vacancy which is filled by a higher shell electron with the emission of an X-ray photon. This provides a trigger for the event and projecting back from the X-ray detection point to the source position provides the time origin for the subsequent recoil time of flight measurement.

However, an additional complication is that the filling of the K shell leaves another higher shell vacancy, and subsequent cascade of atomic electron transitions results in some being ejected from the atom as Auger electrons (typically within $10^{-13}$ -$10^{-15}$s [66]). The latter have significant momentum which has to be included in the mass reconstruction (Fig 3.5b). Thus the K-capture decay mode, although initially giving only an atomic recoil followed by an X-ray, also needs measurement of one or more Auger electrons, but with the advantageous difference that for appropriate choice of initial X-ray shell, the Auger electrons are much lower in energy/momentum compared with the beta decay spectra discussed in §5, and hence in principle easier to measure.

There are 65 K-capture isotopes with half lives between a few days and two years, but many can be eliminated by including only those which decay directly to the nuclear ground state of the preceding element, to avoid additional nuclear gammas contributing to the kinematics or giving additional triggers. Those remaining are shown in Table 6.1, listed as before in order of net event rate per year per atom.

**Table 6.1** K-capture isotopes with no accompanying nuclear gamma, showing possible X-ray trigger energies. Listed in order of triggered decay rate/year/atom (number in MOT maintained constant).

| Z | | A | half life days | Q keV | trigger X-rays KL, KM, KN keV | fraction without gamma | trigger rate per year per atom | MOT |
|---|---|---|---|---|---|---|---|---|
| 55 | Cs | 131 | 9.6 | 352 | 29.8, 33.6, 34.4 | 1.00 | 26.23 | <-- |
| 74 | W | 178 | 22 | 91 | 57.5, 65.2, 67.0 | 1.00 | 11.45 | |
| 38 | Sr | 82 | 26 | 180 | 13.4, 15.0, 15.2 | 1.00 | 9.69 | <-- |
| 24 | Cr | 51 | 28 | 752 | 5.4, 5.9, --- | 0.90 | 8.10 | <-- |
| 77 | Ir | 189 | 13 | 532 | 63.0, 71.4, 73.4 | 0.39 | 7.56 | |
| 18 | Ar | 37 | 35 | 813 | 2.6, 2.8, --- | 1.00 | 7.20 | |
| 4 | Be | 7 | 53 | 861 | 0.045 (Auger) | 0.90 | 4.28 | |
| 33 | As | 73 | 80 | 340 | 9.9, 11.0, 11,1 | 0.90 | 2.83 | |
| 34 | Se | 75 | 120 | 864 | 10.5, 11,7,11.9 | 1.00 | 2.10 | |
| 66 | Dy | 159 | 144 | 366 | 44.5. 50.3, 51.7 | 0.73 | 1.28 | <-- |
| 32 | Ge | 68 | 270 | 107 | 9.2, 10.3, 10.4 | 1.00 | 0.93 | |
| 23 | V | 49 | 338 | 602 | 4.5. 4.9, --- | 1.00 | 0.75 | |
| 73 | Ta | 179 | 653 | 110 | 55.8, 63.1, 65.0 | 1.00 | 0.39 | |
| 26 | Fe | 55 | 986 | 232 | 5.9, 6.4, --- | 1.00 | 0.26 | |
| 79 | Au | 195 | 186 | 227 | 66.8, 75.7, 77.9 | 0.10 | 0.14 | |

As for Table 5.1, those elements which have been successfully trapped in an MOT, listed in Table 5.2, are indicated in the final column of the table. Top of this list is $^{131}$Cs, which is also considered one of the best prospects for achieving an MOT source with $10^8$ -$10^{10}$ atoms, and will therefore be assessed in more detail in the remainder of this section. The special case of $^7$Be, which has particularly simple kinematics, with no X-ray but a single low energy Auger electron trigger, will be discussed in §7.

Electron capture in $^{131}$Cs is predominantly from the K shell, but can also occur from higher shells. Using the theoretically calculated L/K ratio ~ 0.14 ± 0.01 [67] and the M/L ratio ~ 0.22 ± 0.01 [68], gives the capture percentages shown in the top row of Table 6.2. These are consistent with the similar percentages reported for neighboring elements $^{125}$I and $^{127}$Xe [69]. The resulting capture vacancy is then filled by an electron from a higher shell with the emission of an X-ray. The X-ray energy ranges and percentages from L, M, and N shells are shown in the remainder of the table [58].

**Table 6.2** Electron capture percentages for $^{131}$Cs, with ($^{131}$Xe) X-ray energies
And percentage transitions from higher shells

|  | electron captured from shell | | |
|---|---|---|---|
|  | K | L | M |
|  | 85% | 12.5% | 2.5% |
| X-rays from shell | | | |
| L | 69.5% | | |
| M | 13.0% | 10.0% | |
| N | 2.5% | 2.5% | 2.5% |
|  | | | |
| X-ray energy range (keV) | | | |
| L | 29.4 - 29.8 | | |
| M | 33.6 - 33.7 | 3.6 – 4.5 | |
| N | 34.4 - 34.5 | 4.6 – 5.3 | 0.7 – 0.9 |

The X-rays in the 29 – 35 keV range would provide a suitable fast trigger, using either LYSO or BaF$_2$ scintillator followed by multi-anode silicon photomultiplier SiPM for position sensitivity, as discussed in §5 for a higher energy gamma trigger. However, the limited light output and energy resolution requires more detailed consideration in the X-ray case. BaF$_2$ has both a fast decay (0.6-ns, 220nm, 1.4 photons/keV) and slow decay (600ns, 310nm, 9.5 photons/keV) light component. The latter output can be doubled by operating at reduced temperature [70]. The fast component is not increased at reduced temperature, and requires wavelength-shifting. However, the total light output is only marginally sufficient for identifying the specific X-ray lines. An X-ray energy ~30 keV yields ~600 photons, producing, assuming a back reflector and ~40% SiPM quantum efficiency, ~100 photoelectrons (pe) with a resolution ~ ±10 pe ~ 10%, or ± 3 keV. This would be sufficient to distinguish an L X-ray, but insufficient to identify the individual M and N X-rays. However, this can be done by a mathematical procedure, based on the fact that the X-ray lines have accurately known values. Starting from the K-capture mass reconstruction equation (Fig 3.5b):

$$m_\nu^2 = [Q - E_\gamma - E_a - E_N]^2 - [\mathbf{p}_\gamma + \mathbf{p}_{ea} + \mathbf{p}_N]^2 \qquad (6.1)$$

(where the subscript 'a' denotes the total energy and net momentum of one or more Auger electrons) differentiation gives, by analogy with eq.(5.2b), now with Q = 350 keV and X-ray energy 30 keV:

$$d(m_\nu^2) \text{ (keV}^2) \approx 640 \text{ (keV)} \times d\mathbf{p}_\gamma \text{ (keV)} \qquad (6.2)$$

showing that an error ~ 0.5 - 1keV will give a reconstructed value of (mass)$^2$ deviating from zero by ± several hundred keV$^2$. Thus, for each X-ray from M and N shells, it is sufficient for the detector to identify the X-ray energy range 32 – 36 keV, and repeat the reconstruction (in real time) for each of the known M and N X-ray energies, to reject the unreal reconstruction and choose the correct X-ray line.

The emission of an X-ray leaves a vacancy which is filled by an atomic electron usually leading to a cascade resulting in the emission of one or more Auger electrons. Fig 6.1 shows an illustrative example of the emission of two Auger electrons following the emission of a K-M X-ray. The energy of an emitted Auger electron is approximately the energy released by the internal atomic transition minus the binding energy of the shell from which the Auger electron is ejected.

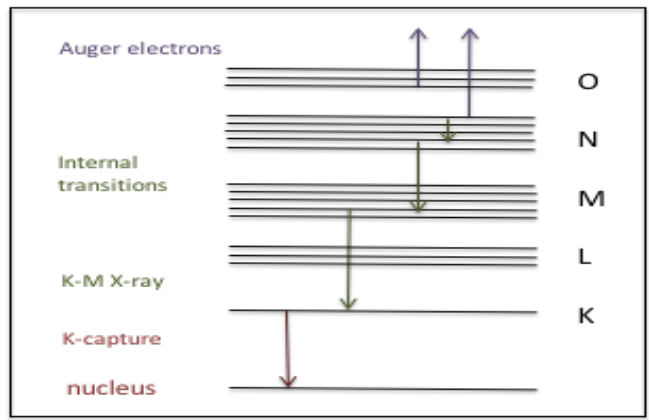

**Fig 6.1** Illustrative example of atomic transitions leading to ejection of two Auger electrons, following K-capture and the emission of a K-M X-ray.

The subshell binding energies of the recoil Xe atom are (in eV),
   L-shell:   5453, 5107, 4786.
   M-shell: 1149, 1002, 941, 689, 676.
   N-shell: 213, 147, 145, 69, 67.
   O-shell:  23, 13, 12,
From this it follows that there are, stemming from any vacancy left by the X-ray, many combinations of internal transitions, leading to a large range of numbers and energies of Auger electrons. However, there are fewer Auger electrons, and lower in energy, resulting from M- or N-shell vacancies [72][73], in particular the N-shell vacancy gives only two principal Auger sequences, type NNO and NOO (standard notation: shell of initial vacancy, shell of electron which fills vacancy, shell of ejected electron) with NOO energies ~ 100-120 eV (single Auger) and NNO energies ~ 20-60 eV (two Augers per decay). The ratio of NOO to NNO Augers is about 25% [72].

Auger electrons stemming from the M shell vacancy have a somewhat larger average energy, ~ 400 eV for types MXY (X,Y, = any higher shell) and ~ 100eV for types MMX [74]. From Table 6.2, restriction to triggering on N-shell X-rays would allow the use of only 2.5% of decays, whereas using both N- and M-shell X-rays would increase this to 15% of decays. We therefore need to design for the collection and time-of-flight measurement of several Auger electrons in the energy range ~ 20 - 400 eV for reconstruction of the majority of decay events that are followed by M- and N-shell X-ray triggers. An important point is that, although these Auger (kinetic) energies are less than 1% of the X-ray trigger energy, they correspond to electron momenta $p_e$ ~ 5 – 20 keV and thus contribute significantly to the mass reconstruction eq. (6.1).

For a more detailed analysis of the decay process, the sequence of events is shown in Fig 6.2.
(a) K-capture in $^{131}$Cs converts it to an excited $^{131}$Xe atom with emission of a neutrino. At this point the recoil and neutrino have equal and opposite momenta.
(b) The K-shell is filled from an outer shell, emitting one of several X-rays, as indicated in Table 6.2. This causes a small change in recoil momentum, depending on the X-ray momentum and (random isotropic) emission angle θ.
(c) Further rearrangement of outer electrons results in one or more Auger electrons to be emitted, again altering the recoil direction by an amount that depends on the random (isotropic) angle of the vector sum of the Auger electrons, denoted by α, relative to the final recoil direction. There may also be one or two 'shake-off' electrons of about 10eV [75][76].

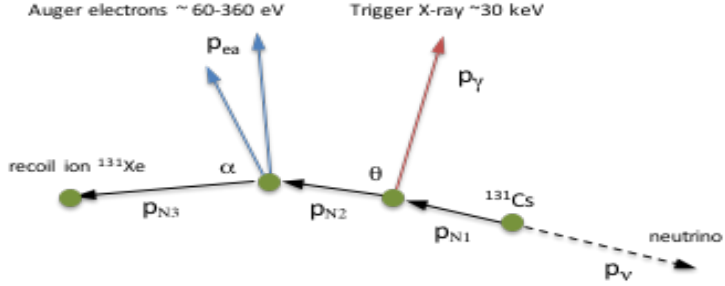

**Fig 6.2** Sequence of events and notation for $^{131}$Cs decay following K-capture

Both X-ray and Auger electrons are emitted within < 1 ps, so the X-ray detection (allowing for the transit time from source to detector) is therefore an adequate trigger for sub-ns time of flight measurement precision. The kinematics of the above processes is given by the following formulae:

(a) $p_{N1}^2 = (Q^2 - m_\nu^2)/(1+Q/M)$ with sufficient approximation, from $p_\nu = p_{N1}$
where Q = 352 keV (difference of $^{131}$Cs and $^{131}$Xe atomic ground states) (6.3)

(b) $p_{N1}^2 = p_{N2}^2 + p_\gamma^2 + 2 p_\gamma p_{N2} \cos\theta$ (6.4)

(c) $p_{N2}^2 = p_{N3}^2 + p_{ea}^2 + 2 p_{ea} p_{N3} \cos\alpha$ (6.5)

(d) the neutrino mass reconstruction formula given above as eq.(6.1).

It can be seen from Fig 6.2 that these equations also cover the fully three dimensional kinematics, since the X-ray vector can be rotated 4π in azimuth around the $p_{N2}$ vector, and similarly the combined **p**$_{ea}$ vector can be rotated 4π in azimuth around the $p_{N2}$ vector, without affecting equations (6.3) – (6.5). Thus the two dimensional form of Fig 6.2 covers all isotropic directions of emission of X-ray and Auger emission, provided the latter are combined into a single resultant **p**$_{ea}$.

It is also important to note that by expansion of eq.(6.1) in the presence of measurement errors, it follows that at zero mass changes in the quantity (mass)$^2$ are linearly proportional to any individual error. Thus insofar as the measurement errors are Gaussian, it is the distribution of (mass)$^2$ that will be Gaussian. This is also the case for non-zero masses at least into the several hundred keV region, and can be assumed in estimating the overlap of the tail of this distribution with any non-zero mass signal. However, in this type of experiment individual errors are likely to be truncated due to the finite extent of detector components, dropping to zero usually in 3 – 4 sigma, which would be beneficial in searching for a separated (mass)$^2$ signal.

To determine the effect of measurement errors the above sequence of equations were set up as a numerical simulation, first calculating $p_{N3}$ with all errors set to zero, then reversing the process to calculate eq.(6.1) in the presence of Gaussian timing errors for $p_{N3}$ and $p_{ea}$ and Gaussian errors in θ or α (using $d\cos\theta = \sin\theta\, d\theta$, $d\cos\alpha = \sin\alpha\, d\alpha$, with the sin factors averaged for isotropic emission). The effect of individual measurement errors on the reconstructed $dm^2$ is summarized in Figs 6.3 and 6.4, assuming an X-ray energy of 30 keV, and Auger energies of 60eV, 120 eV, 360 eV. The time-of-flight path length L is included in the axis of Fig 6.3. The dashed line in Fig 6.4 for an X-ray of 4 keV corresponds to the possibility of using L-capture events, as discussed below in §7. Angle errors would be basically detection position resolution divided by path length, but require additional simulations if electric fields are used to improve solid angle capture of the ion and electrons. An additional factor is the likelihood of several possible charge states arising from emission of alternative numbers of Auger electrons, which can be identified from the transit time and allowed for in the time of flight momentum calculation.

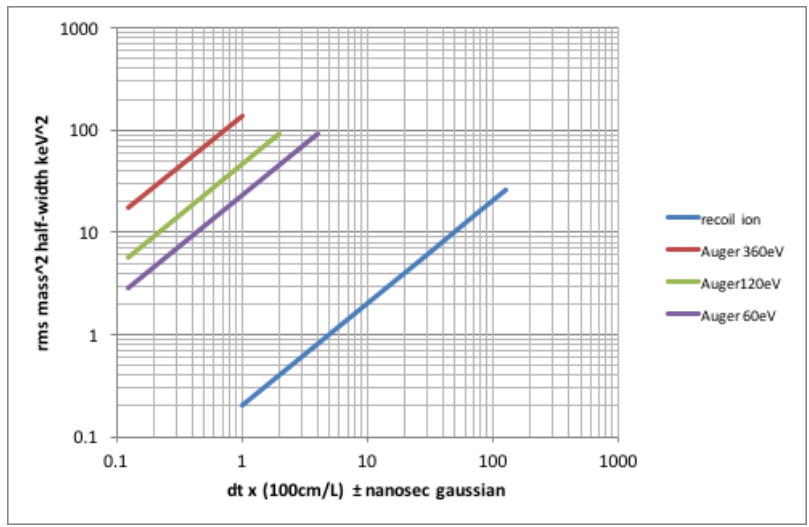

Fig 6.3  Half-width of neutrino (mass)$^2$ distribution versus rms time-of-flight errors for $^{131}$Cs.
   The results are for an assumed time of flight path L = 100cm, the effect of
   other path lengths being included by the factor 100cm/L in the axis parameter.

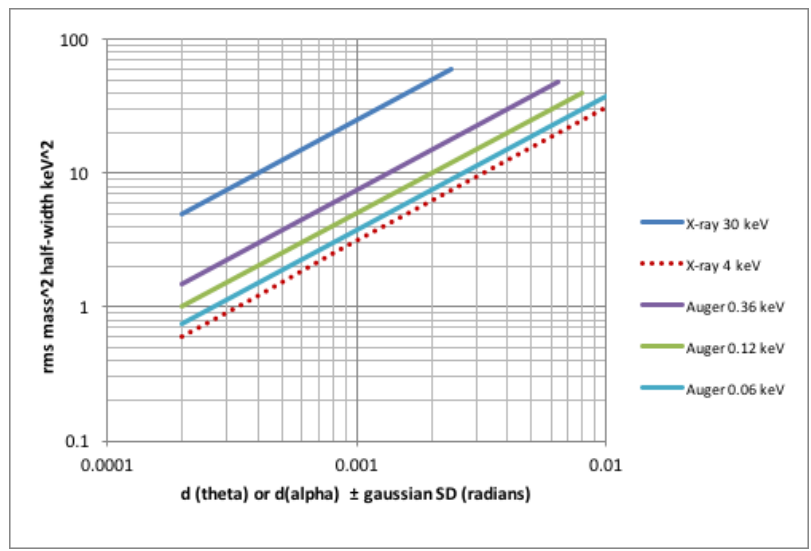

**Fig 6.4**  Half-width of neutrino (mass)$^2$ distribution versus emission angle errors for $^{131}$Cs.
   (with $d(\cos\theta,\alpha) = \sin\theta,\alpha\, d(\theta,\alpha)$ and sin factor averaged for isotropic emission)

## 7 Illustrative experimental configuration

For an experimental sensitivity to sterile neutrino masses down to the region 7-10 keV, or 50 -100 keV$^2$, a reconstruction width $d(m^2) < \pm$ 10-20 keV$^2$ is required, or less depending on the number of processed events per year needed to reach a given coupling level. Allowing for the fact that there are four measurement errors to be combined in quadrature, the individual errors in Figs 6.3, 6.4, should be halved. For the relatively slowly moving recoil ion, a 50cm path length is sufficient, but the faster Auger electrons need at least a 2m path, if the M-shell Augers are to be included. For the X-ray and Auger electrons, requiring a (halved) angular precision 0.0002 – 0.0005 radians, a 0.2 – 0.4mm position resolution requires a 1-2m path length.

This leads to an illustrative configuration of the form and size shown schematically in Fig 7.1, which proposes a conventional MCP detector for the recoil ions, and a large solid angle of a new design of 'pho-switch' detector, utilizing BaF$_2$ scintillator for X-rays and plastic scintillator for pre-accelerated (2kV) electrons, the photons from both being observed by a multi-anode SiPM array, coated with wavelength shifter to respond to the fast BaF$_2$ component. The back of the plastic scintillator could be coated with a 300 Å reflective metallic layer to redirect 90% [77] of back-emitted light into the forward direction while allowing the pre-accelerated electrons into the plastic scintillator.

A finer position resolution (< 40 microns) is obtainable for the Auger electron detection using an MCP array plus guiding electric fields, but using separate detectors for electrons and X-ray reduces the solid angle collection for each, with a reduction in trigger efficiency and the collection of multiple Augers.

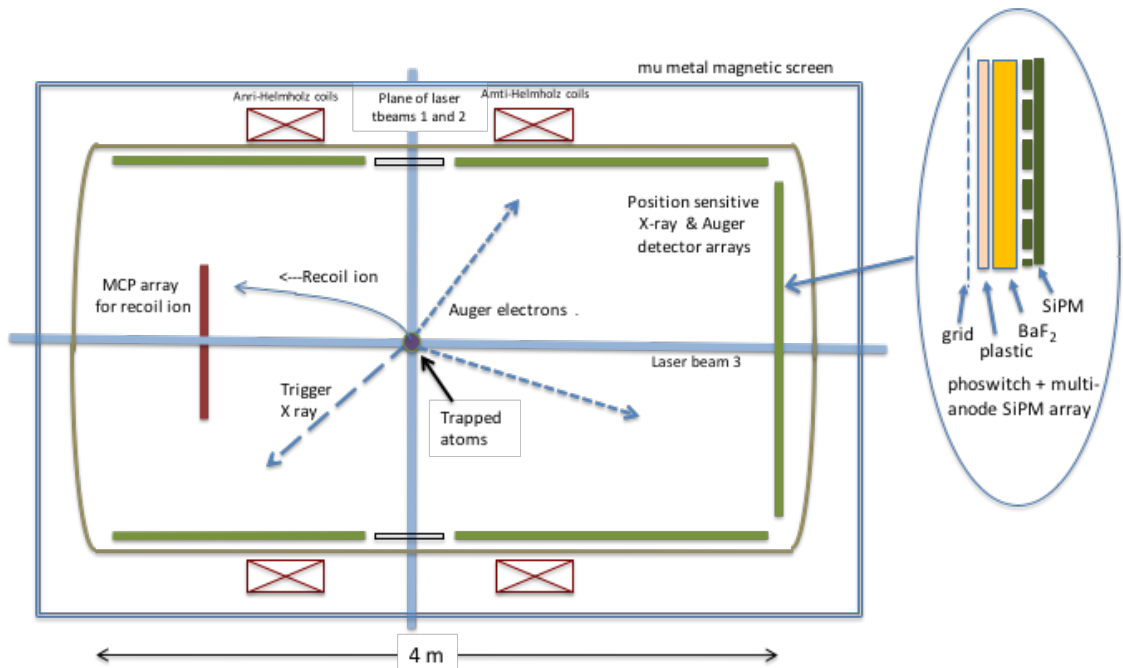

**Fig 7.1** Illustrative arrangement for measurement of decay products from K-capture, based on $^{131}$Cs atoms trapped at the intersection of three retro-reflected laser beams plus an anti-Helmholtz magnetic field (MOT configuration). Recoiling $^{131}$Xe atoms are directed by electric fields on to a multichannel plate, while both X-rays and Auger electrons are registered by a surrounding large solid angle pho-switch detection array. The latter, shown in the inset enlargement, detects 30 keV X-rays with a 1mm thick BaF$_2$ scintillator coupled to a wavelength shifter plus multi-anode silicon photomultiplier for X-ray detection, with electrons detected by acceleration on to a preceding thin plastic scintillator layer.

A system on this scale, using state-or-the-art sub-ns time and sub-mm position resolution, would be capable of separating reconstructed neutrinos of (mass)$^2$ down to the region 50 - 100 keV$^2$, and hence sensitive to the predicted sterile neutrino mass range predicted as a solution to the dark matter problem, in particular the possible observation (Fig 2.4) of an X-ray signal that may correspond to the decay of a 7 keV mass particle.   However, the coupling sensitivity is largely independent of the measurement precision, depending on the number of events that can be processed in, for example, a one-year running time.  Since the decay process is isotropic, this depends on the decay rate, the fraction of decays with accepted triggers, the solid angle capture efficiency of all decay products, and the detection efficiency for each type of particle.  The estimated numbers for these processes are summarized in Table 7.1, for four different choices of capture shell and X-ray shell:

*Column 1:*   K-capture with X-ray triggers from N-shell only.  The Auger electrons then arise only from the N and O shells, with an estimated average number ~ 1.5 per event.

*Column 2:*  K-capture with X-ray triggers from both M and N shells. The Auger electrons arise from processes in M, N, and O shells, with an estimated average number ~ 2.7 per event.

*Column 3 & 4:* These apply to the possibility of utilizing also the electron captures from the L shell which, although only increasing the number of accepted decays by 15%, in fact would nearly double the X-ray triggers from the M and N shells (Table 6.2), and thus double the sterile coupling sensitivity.

The summed contributions from the different capture and X-ray shells are summarized in Table 7.2. A possible problem of using L-capture events, as seen from Table 6.2, is that the M and N  X-ray energies are in the lower range 4 - 5 keV. Using the $BaF_2$ resolution estimates for 30 kev X-rays in §6, the energy resolution becomes ~ 20 – 25% at 4 keV, or  ~ ± 1 keV, again exceeding the spacing of the L X-ray lines, but still with the possibility of using the comparative reconstruction procedure described in §6 to choose the correct tabulated line.  It thus seems possible in principle to utilize the L-capture events, but needs some additional numerical study (along with, eventually, practical confirmation). If confirmed, the sensitivity figures in the final column of Table 7.2 should be feasible, provided that the large MOT filling can be achieved and that the resulting atom cloud size increase, eq.(4.1), can be adequately compensated by the time-focusing technique, eq.(4.4).

**Table 7.1**   Estimates of coupling sensitivity after 365 days active running, for four different choices of initial capture shell (K or L) and X-ray shell (M or N), as discussed in the text., with an MOT filling of $10^{11}$ Cs atoms. The electron detection efficiencies take account of the average number of Augers by applying the latter as an exponent.

| capture shell | K-capture | K-capture | L-capture | L-capture |
|---|---|---|---|---|
| X-rays used: | N-shell | M-shell | N-shell | M-shell |
| Augers: initial vacancy | N+O | M+N+O | N+O | M+N+O |
| Augers: average number per trigger | 1.5 | 3 | 1.5 | 3 |
|  |  |  |  |  |
| fraction of decays | 0.025 | 0.13 | 0.025 | 0.1 |
| useable decays/y/atom | 0.66 | 3.42 | 0.66 | 2.63 |
| ion capture fraction (2 sides) | 0.8 | 0.8 | 0.8 | 0.8 |
| ion detection efficiency | 0.7 | 0.7 | 0.7 | 0.7 |
| electron capture fraction | $0.9^{1.5}$ | $0.9^3$ | $0.9^{1.5}$ | $0.9^3$ |
| electron detection efficiency | $0.7^{1.5}$ | $0.7^3$ | $0.7^{1.5}$ | $0.7^3$ |
| X-ray capture fraction | 0.9 | 0.9 | 0.9 | 0.9 |
| X-ray detection efficiency | 0.5 | 0.5 | 0.5 | 0.5 |
| overall detection efficiency | 0.126 | 0.063 | 0.126 | 0.063 |
| net events/year/source atom | 0.08 | 0.215 | 0.08 | 0.166 |
| atoms in MOT | $1.10^{11}$ | $1.10^{11}$ | $1.10^{11}$ | $1.10^{11}$ |
| processed events per year | $0.8.10^{10}$ | $2.1.10^{10}$ | $0.8.10^{10}$ | $1.7.10^{10}$ |

Table 7.2  Summed contributions for increasing capture and shell contributions, and consequent coupling 'sensitivity', defined as the reciprocal of number of events processed and, in the event of no significant signal, to be multiplied by 2.3 to obtain the 90% Poisson limit on zero. A significant number of signal events will require a coupling a factor ~ 5- 10 higher than the limiting sensitivity. The figures in red would reach the hypothetical observational estimate in Fig 2.4.

| capture & X-ray shells | K, N | K, M+N | K+L, N | K+L, M+N |
|---|---|---|---|---|
| Total events/y | $0.8.10^{10}$ | $2.9.10^{10}$ | $3.7.10^{10}$ | $5.4.10^{10}$ |
| sensitivity for $10^{11}$ atoms in MOT | $1.2.10^{-10}$ | $3.4.10^{-11}$ | $2.7.10^{-11}$ | $1.9.10^{-11}$ |

## 8  Event processing and identification

To achieve sensitivity to coupling levels below $10^{-10}$, should this be necessary to confirm the sterile neutrino interpretation of the astronomical observation (Fig 2.4) an event processing rate ~ $10^3 - 10^4$ /s would be needed.  An accepted event would consist of a 32-36 keV X-ray trigger, followed by a cluster of Auger events after ~ 200 ns, and completed by a recoil ion signal after ~200 µs (for a singly charged ion, or other computable times for multiply charged ions).  Applying time windows would largely reject the larger number of triggers not followed by both Auger and ion signals. Fig 8.1 illustrates qualitatively the possible time sequence of successive events, for the extreme case of an MOT with $10^{12}$ trapped $^{131}$Cs atoms.  Fig 8.2 shows an alternative of overlapping events to allow for faster processing with many sequences rejected as missing some signals through detection efficiency.

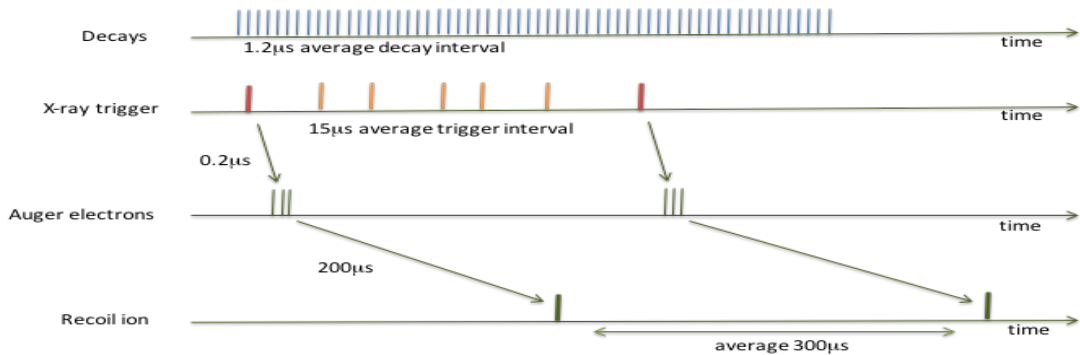

Fig 8.1  Illustrative timelines for accepted decay events from $^{131}$Cs *(timelines not to scale)* awaiting completion of the event before starting a new event with the next X-ray trigger.

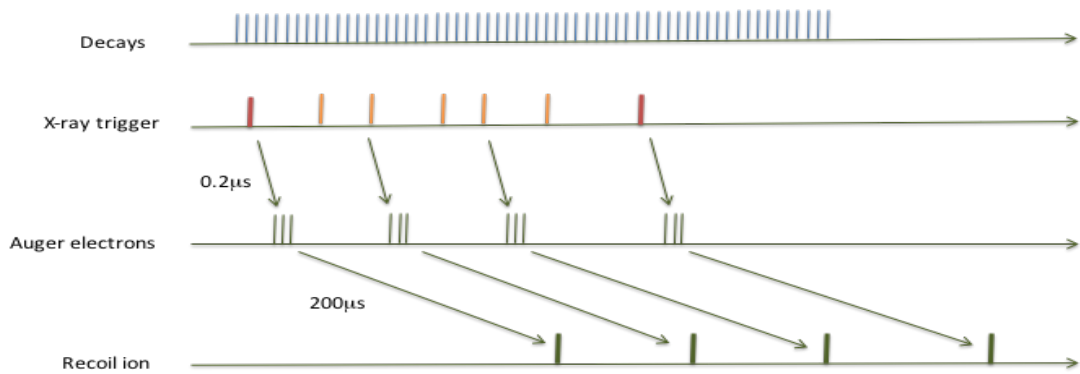

Fig 8.2  Alternative of overlapping timelines for faster event processing, beginning a new triggered event before completion of the preceding event, providing an early start on the next event if the preceding event does not complete.

## 9  The special case of isotope $^7$Be

The K-capture isotopes all have the complication of emitting several Auger electrons, with the need to ensure that, for each accepted event, all are collected along with their momentum and direction, in order to achieve a complete reconstruction of the neutrino mass. The need to minimize their number, restriction to N-shell X-rays, or at most N + M shell X-rays, together with collection and detection efficiency factors for each Auger electron, considerably reduces the overall collection efficiency for completed events (Table 7.1).  The single exception to this in Table 6.1 is the isotope $^7$Be for which 90% of decays are direct to the $^7$Li ground state, with no X-ray emitted but the K-capture vacancy filled with the release of a single Auger electron of known kinetic energy 45 eV [78] (Fig 9.1).

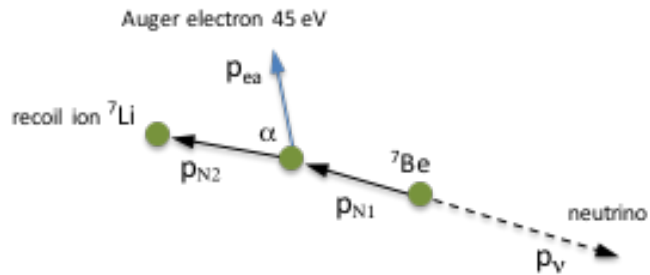

**Fig 9.1**  90% K-capture decay of $^7$Be

This has the following advantages compared with the sequence for other K-capture isotopes (Fig 6.2):

(a) There is no X-ray to be detected, thus avoiding the need for a special detector for this.
(b) A single Auger electron is emitted, with fixed 45 eV energy, providing a time-of-flight trigger.
(c) Collection of the single electron can be achieved with a large solid angle using an electric field.
(d) MCP detection of the trigger electron provides 0.3ns time resolution, and 40 μ position resolution.
(e) The Q value of 860 keV for $^7$Be allows a wider range of sterile neutrino masses to be explored.
(f) The higher Q and lower atomic mass results in a recoil velocity 40 times higher, and kinetic energy 100 times higher, than for Cs, making the source temperature ($\propto Mv^2$) correspondingly less critical.

The higher recoil speed necessitates a longer time-of-flight path than in the case of $^{131}$Cs, to achieve the same momentum and mass precision. For a combined timing error 0.3ns (i.e. ± 0.2ns at start and finish), Fig 9.2 shows the mass$^2$ error as (left) a function of recoil path length L, and (right) a function of the angle error (between electron and atom recoil) for a fixed L = 200cm and timing error 0.3 ns.

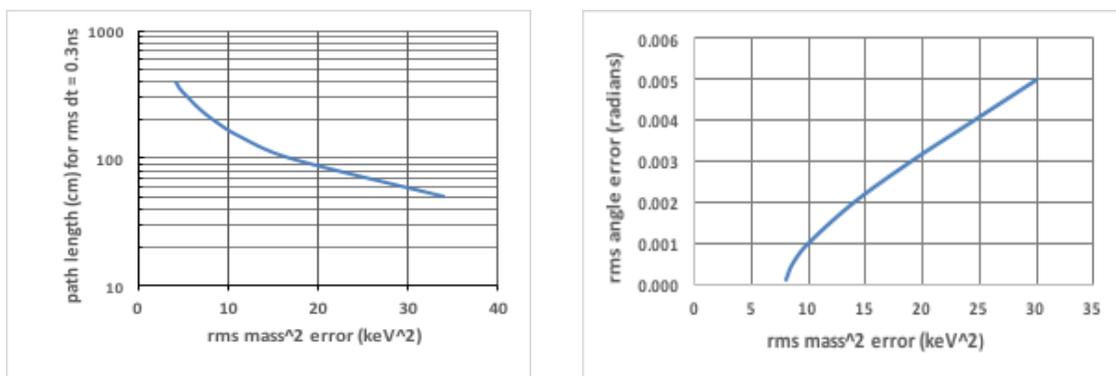

**Fig 9.2**  Calculations for $^7$Be K-capture:
 (left)  Time-of-flight path length versus rms mass$^2$ width, for an assumed 0.3ns rms timing error
 (right) rms angle precision versus rms mass$^2$ width, for fixed path length 200cm and dt = 0.3ns.

The longer path length for the atom recoil would require a combination of axial magnetic and electric fields to capture a large range of emission angles. Typically 450 gauss and 25 V/cm are required to focus emission angles up to 70 degrees in one hemisphere on to a 20cm diameter MCP array at a distance 200cm. Lower fields are required to capture the 45 eV Auger electrons, typically 5 gauss and 10 V/cm, to focus electron emission angles up to 80 degrees (in the opposite hemisphere) on to an MCP array of diameter 20cm placed at a distance ~ 60cm from the source. The two requirements could be matched by means of a magnetic field configuration illustrated in Fig 9.3, where a short reverse magnetic field coil is placed to the right of the source to create a rapid fall in the fringe field of the larger solenoid. Note that although the field distribution is non-uniform there remains a one-to-one correspondence between each detection point on the MCP and the polar and azimuthal emission angles from the source, this being established by computer simulation, together with final calibration via the reconstruction of the majority of events at zero neutrino mass.

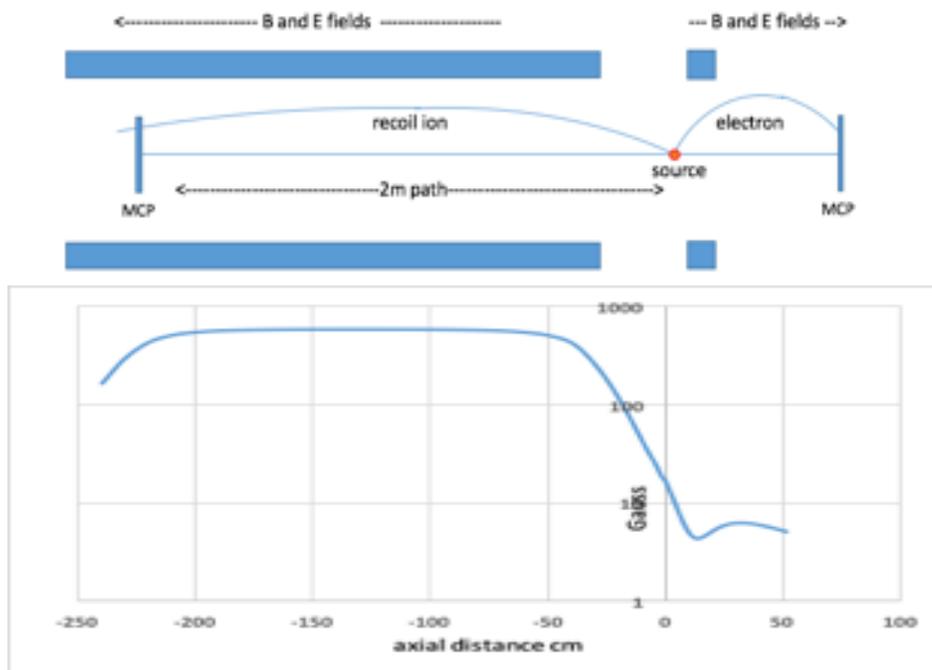

**Fig 9.3** Illustrative axial magnetic and electric field configuration for collection of recoil ions and Auger electrons from $^7$Be K-capture. The fields on the left are 450 Gauss and 25 V/cm, and on the right are 5 Gauss and 10V/cm.

Table 6.1 shows that for $^7$Be the trigger rate per year is a factor ~ 6 lower than for $^{131}$Cs, but this is offset by the above advantages for the efficiency and precision of the experiment, and $^7$Be would be the strongly favored choice for a sterile neutrino search were it not for the fact that at the time of writing it cannot be trapped efficiently in an MOT or a high density beam, and there are further problems arising from positive ion production [54]. Nevertheless, the above advantages provide a strong incentive for the development of a method for making this into a high density source or beam, perhaps via an extension of the 'buffer cell' technique [59] which has been used to produce beams of some previously intractable atoms, compounds, and radicals, and thus suggests an optimistic prospect for the future use of trapped Be in this application.

## 10 Backgrounds

Fig 10.1 indicates the most likely background sources as follows (not in order of significance):
   (a) Cosmic ray muons
   (b) Residual gas scatters by Auger electrons during flight to detectors
   (c) Radioactive impurities in walls and component materials
   (d) Atoms escaping from the trap, or lost during loading, adhering to walls and detector surfaces.
   (e) Radiative atomic transitions, radiative beta decay, or radiative K-capture

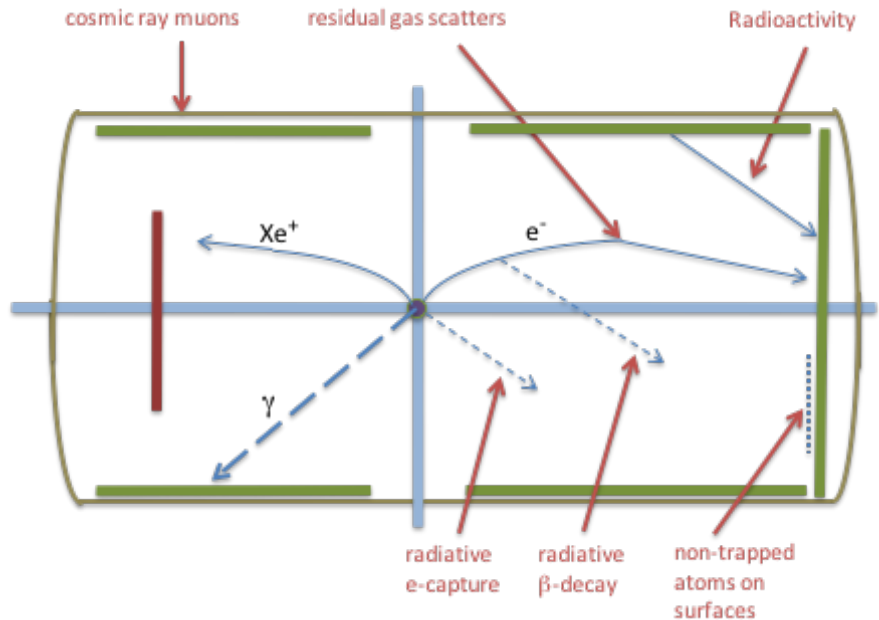

**Fig 10.1**   Simplified version of Fig 7.1 indicating the listed background effects (a) – (e)

A more detailed discussion of each is as follows:

   (a) Cosmic ray muons (assuming the experiment to be located at the Earth's surface) would give a rate (100/m$^2$/s/sr) of signals in individual components of the scintillator array, but these would be predominantly at the MeV-level (for example in 1mm BaF$_2$ scintillator plates), and can be rejected in real time by an energy cut, hence not contaminating the accepted events. The resulting dead time (from the ~ 2μs recovery time of the BaF$_2$) would be < 1%, but most of this would be concurrent with (and would not affect) the timing of the recoil ion to the MCP. Although a surrounding anti-coincidence muon detector could be designed, it is unlikely to be needed, since the scintillator array provides its own anti-coincidence.

   (b) Scattering from residual gas molecules can be minimized with a cryopumped ultra high vacuum, (already a basic requirement to maximize MOT lifetime after loading with Cs atoms) but some scattering events may occur, changing the direction of the decay products. Background from these can be rejected by the previously-mentioned fact that each of the mass$^2$ reconstruction equations (5.1) or (6.1) is the difference between two large numbers ~ 10$^6$ keV$^2$. It is this that, on the one hand, gives the challenge of measuring these numbers with sufficient precision, but on the other hand also provides a significant advantage in rejecting spurious events. In particular, a residual gas scatter would result in an incorrect value for the magnitude and direction of the electron momentum. The resulting large difference, sometimes then negative, between the two terms of Eq. (5.1) or (6.1), would be sufficient to allow rejection of these events.

(c) Radioactive impurities would be reduced to a minimum in the selection of cryostat, source and detector materials, but residual concentrations would emit, electrons and gammas, giving random signals in detectors. These would largely be rejected by imposing time windows on the sequence of signals that constitute an acceptable event. After an X-ray trigger of the required M or N shell energy, a time window ~ 100-200 ns would be allowed for the arrival and detection of the Auger electrons, followed by a second time window ~ 0.5 – 1.5 ms later for the detection of the single recoil ion at the MCP (this range allowing for all sterile neutrino masses up to 300 keV). Only the detection of Auger and ion signals within these time intervals would constitute an acceptable event, leaving a large fraction > 99.9% of the time (see Fig 7.1) in which radioactive background signals would be automatically ignored.  Moreover, any arriving within the correct time window would simulate an additional Auger electron for which the complete event would then not reconstruct consistently. In support of this, it should be noted that this background has not been reported as a problem in many published COLTRIMS experiments [45][46].

(d) Atoms are lost into the vacuum chamber during loading, and from the MOT due to its finite trapping lifetime, which is of order ($10^{-8}$/p Torr) seconds [48]. Thus with ultra-high cryopumped vacuum, with pressure p ~ $10^{-12}$ - $10^{-13}$ Torr, the lifetime could be $10^4$-$10^5$ seconds. The lost atoms then move in straight lines to various surfaces, including the cryostat, the MCP and the large area X-ray/electron detector, where they decay with a 9.7d half-life. Thus in equilibrium (loss rate from the MOT = decay rate) there would be a number of atoms, on surfaces in the system, exceeding those in the trap by a factor ~ 10-100, and hence, once this equilibrium is reached, decaying at a rate  ~ 10-100 times those in the MOT.  However, although these produce both X-rays and Augers, these will either not both be detected or not within the correct time windows.  Specifically, for atoms on either (i) the detector surface or (ii) on its pre-acceleration grid, the Augers will be confined to that detector, and thus too close in time (i.e. within a few ns) to the X-ray signal (travelling at velocity c).  Finally, (iii) for atoms on the interior surface of the chamber, although the emitted X-ray may reach a detector, the electrons can be confined to the surface by lining it with a grid or alternating potentials [79] and will not reach any detector. Thus decays of escaping atoms cannot be mistaken for real events, but nevertheless need to be kept to < 10 times the source decay rate, to avoid pile-up or else too large a dead time from the rejection of these background signals at the highest planned event-processing rates.

(e) There are several processes from which additional photons may be emitted:
(i) In the atomic transitions emitting Auger electrons, the energy can be released alternatively as low energy X-rays. Published calculations list a low relative probability for these, below $10^{-3}$ for the M shell and below $10^{-4}$ for the N shell [73]. These X-rays would have the same low energy as the corresponding Auger transition, and hence not detected, giving unreconstructed events that would be rejected in real time by having incomplete energy collection.
(ii) For beta decay, a radiative diagram exists, in which the free electron emits a photon with a continuous spectrum up to the beta end point energy. This has been calculated for Tritium decay by Bezrukov [9].  An undetected photon of energy $E_{ph}$ in eq.(5.1) would mimic a non-zero neutrino (mass)$^2$ ~ 2 $E_{ph}$ [Q -$E_e$] keV$^2$ (Q =18.6 keV for Tritium) averaged over emission angles relative to the beta electron, with a probability ~ $10^{-5}$-$10^{-6}$ /keV in the mass region 4 – 12 keV. This would appear to make it more difficult to detect a sterile neutrino coupling below this value unless these events can be rejected by detecting the coincident radiated photon.

(iii) In the case of electron-capture there is no free electron to radiate a photon, but radiative capture nevertheless occurs, and the resulting photon spectrum has been studied both theoretically and experimentally [80][81][82][83]. The differential spectrum depends strongly on the capture state (S or P). For K-capture (S-state) the spectrum has the form:

$$dN_{ph}/dy = (\alpha/\pi)(Q/m_e)^2 y(1-y)^2 \tag{10.1}$$

where $y = E_{ph}/E_{max}$, $E_{max} = Q$, $\alpha = 1/137$, $m_e = 511$ keV

giving a total probability per K-capture

$$N_{ph} = (\alpha/12\pi)(Q/m_e)^2 \tag{10.2}$$

The typical case of $^{119}$Sb (with Q=590 keV) provides an example of agreement between theory and measurement [82] and similar agreement has been found for a number of other K-capture elements. For $^{131}$Cs, with Q = 362 keV, $dN_{ph}/dE_{ph}$ is shown as curve (a) in Fig 10.2, confirmed in [83] by absolute measurements of the upper branch in [83]. Its potential effect as a background In a sterile neutrino search is that an undetected photon of energy (and momentum) $E_{ph}$ in eq.(6.1) will, after reconstruction, mimic a non-zero neutrino mass $m_{eff}$, its value dependent on both $E_{ph}$ and the emission angle $\delta$ relative to the nuclear recoil. Using an average $\cos \delta \sim 0$ results in a reconstructed effective mass $m_{eff} \approx (2E_{ph}Q)^{0.5}$, or $m_{eff} \approx [4 E_{ph}(Q - E_{ph})]^{0.5}$ for the case $\cos \delta = \pm 1$. Using the former to convert the $E_{ph}$ axis to an effective $m_{eff}$ axis, together with $dm_{eff}/dE_{ph} = [Q/2E]^{0.5}$, transforms the $dN/dE_{ph}$ spectrum, shown in blue, to the required $dN/dm_{eff}$ spectrum, shown in red. For both spectra, N is the number of events per keV per K-capture. Corresponding curves for $^{7}$Be (Q = 860 keV) is shown in Fig 10.3.

Thus, for any value of reconstructed mass, curves (b) give the probability of a spurious event resulting from radiative capture. To magnify the low end of the spectra, they are shown also on log scales in Figs 10.2 and 10.3. For $^{7}$Be, the higher Q value results in about a factor ~ 2 higher $dN/dE_{ph}$ background, but after conversion to an effective reconstructed mass the radiative background, at any specific mass, is about a factor 4 lower than that of $^{131}$Cs. In the mass region $m_s \sim 10$ keV, Figs 10.2 and 10.3 show that this background would be sufficiently low to reach coupling levels $< 10^{-11}$. This rises to $10^{-8}$ at ~ 100 keV, in principle reducible by an order of magnitude if the radiated photon could be detected (in coincidence with the atomic X-ray) with 90% efficiency, enabling rejection of those events.

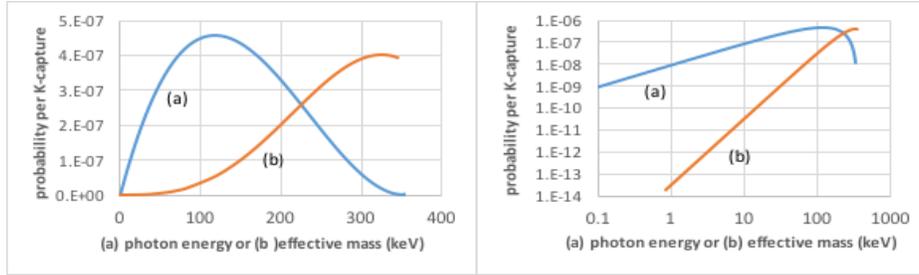

**Fig 10.2** (a) Radiative K-capture spectra for $^{131}$Cs and (b) converted to an apparent reconstructed mass

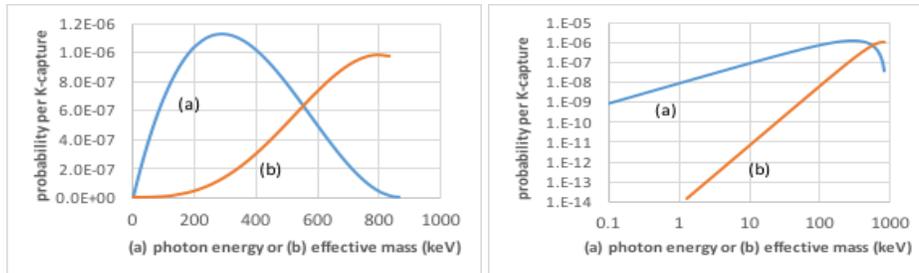

**Fig 10.3** (a) Radiative K-capture for $^{7}$Be and (b) converted to an apparent reconstructed mass

## 11 Other options using K-capture

Polarization of the atomic nucleus, will create an angular asymmetry in the direction of the emitted neutrino relative to the initial spin direction, and hence in the recoil nucleus, whose recoil is initially opposite in direction to that of the neutrino. This has sometimes been taken to imply that for a 99% polarized nucleus the neutrino direction will be almost unidirectional, subject only to thermal fluctuations of the nuclear spin [84][85]. Clearly this would be of value in the mass reconstruction, since one of the directions would be defined in advance, and also allow complete collection of the atomic recoils. However, although the atom and neutrino directions are initially equal and opposite, that direction can nevertheless lie at an angle $\theta$ to the initial polarization direction. The correct angular distribution was calculated by Treiman[86] who derived a dependence of the form $(1 - A \cos \theta)$ where, for a 5/2 -> 3/2 transition, the constant A is equal to the polarization P and, for P ~ 1, the recoils will be predominantly in the same direction as the polarization [87]. While this has some forward/back asymmetry, it may not be sufficient to justify the additional complication of using a polarized source.

Polarization of $^{131}$Cs can be achieved by optical pumping but not while trapped in an MOT. However, if the MOT is turned off rapidly (less than a millisecond), the trapped ions will remain in place for long enough (1-2 ms) for polarization (< 100μs) and data-taking, after which the MOT can be re-established in ~ 2ms [88]. There is thus a repeated ~ 4 ms cycle of alternating trapped and polarized atoms (Fig 10.4). Thus this loses a factor ~ 3 in data-taking time, and for this to be adopted in a sterile neutrino search, any advantage from polarization would need to offset this loss. A similar repetitive sequence without polarization, of simply turning off the MOT while data taking, could be adopted if there is too much perturbation of Auger electron paths by the anti-Helmholtz magnetic field required for the MOT.

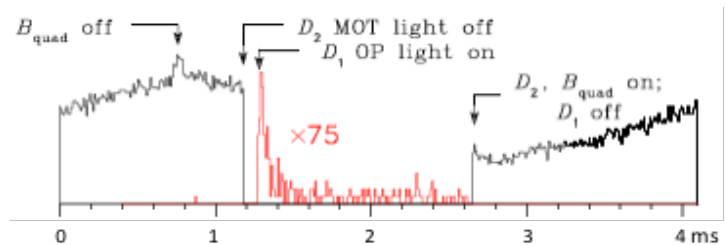

**Fig 10.4** (reproduced from D Melconian Thesis [88]). 4 ms time cycle for polarized $^{37}$K beta decay experiment showing MOT field turn-off, MOT light turn-off, optical pumping, data-taking interval, and restoration of MOT

## 12 Conclusions

Sterile neutrinos in the 10 keV mass region could provide an explanation of the Galactic dark matter. Direct detection appears not to be feasible in the near future, making a strong case for the development of laboratory searches. This detailed study shows that such searches are feasible using full reconstruction of beta decay or K-capture decay events, from neutral atoms suspended in a magneto-optical trap. This provides the only method proposed so far for isolating and observing individual sterile neutrino events as a keV-range mass population, separated from the standard neutrinos close to zero mass. The survey of suitable isotopes in this paper suggests that precise reconstruction of K-capture events appears more feasible than those emitting a beta spectrum. For the most favorable cases $^7$Be and $^{131}$Cs the currently attainable (sub-ns) time-of-flight resolution provides sufficient momentum precision to resolve keV-range neutrino masses in a laboratory-scale experiment. Of these two isotopes, $^{131}$Cs can be trapped in an MOT in large numbers, but presents a more difficult measurement problem because of the need to collect and measure, in addition to the nuclear recoil, an X-ray and several low energy Auger electrons. In contrast, $^7$Be has a much simpler decay process which would make it the best choice from a measurement viewpoint, but cannot yet be efficiently loaded into an MOT or high density beam.

For any choice of isotope, the more difficult prospect is the processing of a sufficient number of events per year to reach low coupling levels.  To make an Improvement on existing coupling limits (~ $10^{-3}$) requires only ~ $10^6$ decays per year, but to reach the coupling level < ~ $10^{-10}$, suggested by a recent unidentified astronomical X-ray signal (Fig 2.4), may require processing of ~ $10^{12}$ decays per year, depending on the efficiency of collection and detection. The latter efficiencies would be much higher (perhaps by two orders of magnitude) for the very simple $^7$Be decay, than in the case of $^{131}$Cs decay with an X-ray and multiple Augers, giving a strong incentive for the future development of some technique to create a compact suspended source of Be atoms.  A further possibility for increasing the event processing rate would be to design a more complex geometry containing multiple sources, or to make, for example, multiple replicas of a single-source apparatus, which would be justified given any hint of a sterile neutrino signal in one system.

A number of background effects have been discussed, most of which can be rejected as not within the correct time windows for real events, or can be rejected by the reconstruction mathematics, which depends on the difference between two large numbers, and hence sensitive to any perturbations arising, for example, from scattering in the vacuum chamber. The one background which cannot be rejected in this way - the photons from radiative K-capture, which on reconstruction would mimic a non-zero neutrino mass, appears to be sufficiently small in the 10 keV mass region to allow couplings below $10^{-10}$ to be reached.  There is thus good reason to believe that the principle of full four-momentum reconstruction in K-capture processes can become a leading method of searching for sterile neutrinos with masses down to the 5 – 10 keV level, and including higher masses up to the Q value of the chosen isotope (~ 300 keV for $^{137}$Cs or ~ 800 keV for $^7$Be).

**Acknowledgements**
The author acknowledges valuable discussions over a period of several years with Eric Hudson (UCLA), Hanguo Wang (UCLA), John Behr (Triumf), Jeff Martoff (Temple), George Fuller (UCSD), Evan Grohs (Michigan), Petr Vogel, (CalTech), Dan Melconian (Texas A&M), and Ottmar Jagutzki (Frankfurt).  This paper covers preliminary studies carried out between 2011 and 2015, subsequently continuing as an ongoing experimental collaboration with Jeff Martoff (P.I.,Temple), Eric Hudson (UCLA), Paul Hamilton (UCLA), Hanguo Wang (UCLA), Andrew Renshaw (Houston).